\documentclass[fleqn,usenatbib]{mnras}

\usepackage{newtxtext,newtxmath}

\usepackage[T1]{fontenc}
\usepackage{ae,aecompl}
\usepackage[dvipsnames]{xcolor}
\usepackage[toc,page]{appendix}

\usepackage{graphicx}	% Including figure files
\usepackage{amsmath}	% Advanced maths commands
\DeclareUnicodeCharacter{2212}{\textendash}

\title[Eclipse Mapping of EXO 0748$-$676]{Eclipse Mapping of EXO 0748$-$676: Evidence for a Massive Neutron Star}
\author[A. H. Knight et al.]{
Amy H. Knight,$^{1}$\thanks{E-mail: amy.knight@physics.ox.ac.uk}
Adam Ingram,$^{1}$
Matthew Middleton $^{2}$
and Jeremy Drake $^{3}$ 
\\
$^{1}$Department of Physics, Astrophysics, University of Oxford, Denys Wilkinson Building, Keble Road, Oxford, OX1 3RH, UK\\
$^{2}$School of Physics and Astronomy, University of Southampton, Highfield, Southampton, SO17 1BJ, UK\\
$^{3}$ Center for Astrophysics | Harvard \& Smithsonian, 60 Garden Street, Cambridge, 02138 MA, USA 
}

\date{Accepted XXX. Received YYY; in original form ZZZ}

\pubyear{2021}

\begin{document}
\label{firstpage}
\pagerange{\pageref{firstpage}--\pageref{lastpage}}
\maketitle

\begin{abstract}
%\cmtak{I've overhauled the abstract. I'm very aware of the 250 word limit and can't get everything the referee wants in here within the word limit without removing the rest of the information in the abstract. I think the below is a good compromise. I've had to remove the sentence about favouring hard equations of state and I've shortened some of the earlier sentences.}
Determining the maximum possible neutron star (NS) mass places limits on the equation of state (EoS) of ultra-dense matter. The mass of NSs in low mass X-ray binaries can be determined from the binary mass function, providing independent constraints are placed on both the binary inclination and mass ratio. In eclipsing systems, they relate via the totality duration. EXO 0748$-$676 is an eclipsing NS low mass X-ray binary with a binary mass function estimated using stellar emission lines from the irradiated face of the companion. The NS mass is thus known as a function of mass ratio. Here we model the X-ray eclipses in several energy bands, utilising archival \textit{XMM-Newton} data. We find a narrow region of absorbing material surrounding the companion star is required to explain the energy-dependent eclipses. Therefore, we suggest the companion may be experiencing ablation of its outer layers and that the system could transition into a redback millisecond pulsar. Our fit returns a mass ratio of $q=0.222^{+0.07}_{-0.08}$ and an inclination $i = 76.5 \pm^{1.4}_{1.1}$. Combining these with the previously measured radial velocity of $410 \pm 5$ km/s, derived from Doppler mapping analysis of H$_\alpha$ emission during quiescence, returns a NS mass of $\sim 2~M_\odot$ even if the line originates as far from the NS as physically possible, favouring hard EoS. The inferred mass increases for a more realistic emission point. However, a $\sim 1.4~M_\odot$ canonical NS mass is possible when considering radial velocity values derived from other emission lines observed both during outburst and quiescence.
\end{abstract}

\begin{keywords}
Accretion: Accretion Discs -- Stars: Neutron Stars -- X-rays: Binaries
\end{keywords}

%and thus a neutron star mass of $M_{\rm ns} = 2.02^{+0.29}_{-0.27}~ M_{\odot}$, which is larger than the canonical mass of $1.4~M_{\odot}$ with $> 3~\sigma$ confidence. 
%%%%%%%%%%%%%%%%%%%%%%%%%%%%%%%%%%%%%%%%%%%%%%%%%%

%%%%%%%%%%%%%%%%% BODY OF PAPER %%%%%%%%%%%%%%%%%%

\section{Introduction}
Observing neutron star (NS) low-mass X-ray binaries (LMXBs) provides an opportunity to constrain the equation of state (EoS) of matter at extreme densities and develop binary evolution models \citep{Steiner2010, Providencia2019, PPBinary, Postnov2014}. Since the EoS (the pressure-density relation) uniquely predicts the NS mass-radius relation (\citealt{Lindblom1992}; see Fig 10 of \citealt{Voisin2020} for some up-to-date examples), it can be constrained from measurements of the mass and radius of astrophysical NSs. Mass alone can be highly constraining since each EoS predicts a maximum possible NS mass. Observationally determining the maximum NS mass also informs stellar evolution models \citep{Antoniadis2016, Sukhbold2018, Raithel2018} and the interpretation of gravitational wave observations \citep{Yang2018, Essick2020, Chen_2020}, both of which typically assume that compact objects with mass $< 3~M_\odot$ are NSs, with the remainder being black holes (BHs). For example, GW190425 \citep{Abbott2020a} was classified as the coalescence of two NSs because the mass of the primary was $\sim 2.5 M_{\odot}$, but no electromagnetic signatures were detected to confirm a NS was present. To date, the most massive confirmed NS is PSR J0740+6620 at $M_{\rm ns} \sim 2.1 M_{\odot}$ \citep{Cromartie2019,Fonseca2021,ColemanMiller2021}. The recent detection of GW190814 \citep{Abbott2020}, classified as the merger of a $\sim 2.6 M_{\odot}$ NS and $\sim 5 M_{\odot}$ BH contests this, suggesting that NSs can exist within the so-called \textit{mass-gap} and impacting the inferred nucleonic EoS \citep{Fattoyev2020}. 

There are many methods used to estimate the mass and/or radius of a NS (see \citealt{Miller2013} or \citealt{Ozel2016} for a summary). Waveform modelling of X-ray pulsations (or thermonuclear burst oscillations), from either isolated or accreting NSs, provides constraints on both mass and radius because the waveform is distorted by Doppler shifts from NS rotation and gravitational redshift \citep{VanParadijs1979, Fujimoto1986, Sztajno1987, Poutanen2003, Riley2019}. Alternatively, the NS radius can be estimated by combining the temperature and observed flux of thermal emission from the NS's surface with a distance estimate. Photospheric radius expansion (PRE) bursts from LMXBs enable a mass estimate by assuming that the burst occurs when the NS is accreting at the Eddington luminosity \citep{Damen1990, Lewin1992, Ozel2006, Guver2010}. The mass of NSs in binary systems can be constrained dynamically by measuring Doppler shifts caused by the orbital motion of one of the binary components. This is the most direct method of measuring NS mass as it assumes only Kepler's laws. For pulsating NSs the Doppler shifts can be measured from orbital phase-dependent shifts to the pulse frequency \citep{Chakrabarty1998}, and for quiescent systems, they can be measured from orbital phase-dependent shifts to the wavelength of lines in the spectrum of the companion star (the radial velocity curve; e.g. \citealt{Casares2014}). In both cases the actual observable is the binary mass function, which depends on NS mass, $M_{\rm ns}$, mass ratio, $q=M_{\rm cs}/M_{\rm ns}$ where $M_{\rm cs}$ is the companion star mass, and binary inclination angle, $i$. In the latter case, the binary mass function is given by
\begin{equation}
    f = \frac{P K^3}{2\pi G} = \frac{ M_{\rm ns} \sin^3 i }{ (1+q)^2 },
    \label{eqn:f}
\end{equation}
where $P$ is the orbital period, $G$ is Newton's gravitational constant and $K$ is the semi-amplitude of the radial velocity curve. The degeneracy between mass, mass ratio and inclination is partially broken in eclipsing systems, in which $q$ and $i$ are related via the duration of totality, $t_e$, if the companion star is filling its Roche-Lobe \citep{Horne1985}. The NS's mass in an eclipsing system with a measured binary mass function is therefore known as a function of $q$.

Here we consider the LMXB EXO 0748$-$676, which underwent a $>20$ yr outburst \citep{Parmar1986} during which many eclipses with $t_e \approx 500$ s were observed by \textit{EXOSAT}, \textit{RXTE} and \textit{XMM-Newton}, recurring on the orbital period of $P=3.824$ hrs \citep{Parmar1986, Wolff2009}. Pulsations have never been detected from the source, but the NS spin frequency is likely within a few Hz of the measured $\sim 552$ Hz burst oscillation frequency \citep{Galloway2010}. \cite{Ozel2006} used PRE bursts to estimate the mass and radius of the NS to be $M_{\rm ns} = 2.10 \pm 0.28~M_{\odot}$ and $r_{\rm ns} = 13.8 \pm 1.8$ km respectively, and showed that such values rule out soft equations of state \footnote{Harder/stiffer equations of state are ones whereby pressure increases more steeply with density}. EXO 0748$-$676 entered quiescence in late 2008 (see \citealt{Wolff2011} for a summary), providing the opportunity to confirm this high mass value dynamically. Absorption lines have never been detected from the companion star, but irradiation-driven emission lines have been observed to be modulated on the orbital period \citep{Pearson2006, MD2009, Bassa2009}. The radial velocity curve of such emission lines can only provide a lower limit on the mass function because they originate from somewhere between the companion and binary system centres of mass \citep{MunozDarias2005}. The resulting lower limit on the mass is $M_{\rm ns} > 1.27~M_\odot$ \citep{Bassa2009}. Thus, further constraints on $q$ are required to verify the mass value presented by \citet{Ozel2006}.

\citet{Ratti2012} attempted to constrain $q$ by measuring the width of the phase-resolved companion star emission lines. Under the assumptions that the star is tidally locked, fills its Roche-Lobe and that the width results entirely from rotational broadening, $q$ can be derived from the line width and radial velocity \citep{Wade1988}. However, they found that the lines were broader than expected, requiring a $>3.5~M_\odot$ NS if rotational broadening dominates. They instead concluded that extra broadening was likely contributed by a stellar outflow, driven by a pulsar wind and/or X-ray heating. The same scenario can also explain the lack of observed emission lines from an accretion disk one year into quiescence, since the evaporation of material during the extended periods of outburst would result in the lack of disk material and indeed emission lines \citep{Ratti2012}. Note that such a non-detection of disk emission lines is incredibly unusual for LMXBs \citep{Marsh1994}. A similar X-ray induced evaporative wind was considered for EXO 0748$-$676 by \citet{Parmar1991} in order to explain the heavily extended ingress and egress durations and their drastic variability. The authors suggested that the evaporative wind was required to sufficiently extend the ingress and egress durations since the atmospheric scale height of the companion should otherwise be $\sim 100$ km. Additionally, the pulsar wind hypothesis is supported by the detection of a broad C IV emission line by \citet{Parikh2021}, who draw similarities between their quiescent observations of EXO 0748$-$676 and the known transitional redback pulsar, PSR J1023+0038 in its rotation powered state. These scenarios are reminiscent of so-called spider pulsars, which are millisecond radio pulsars with an under-massive companion star that is in the process of being ablated by an ionising pulsar wind. They are further subdivided into redbacks where $0.1 M_{\odot} < M_{\rm{cs}} \leq 0.5 M_{\odot}$ and black widows \citep{Fruchter1988} where $M_{\rm{cs}} \leq 0.1 M_{\odot}$. Spider pulsars represent a key stage of pulsar evolution under the paradigm that isolated millisecond radio pulsars were spun-up by accretion before completely consuming their donor star \citep{Rad1982, Alpar1982}. Transitional millisecond pulsars (tMSP), many of which are classified as redback pulsars \citep{Linares2014}, switch between accretion powered X-ray pulsations and rotation-powered radio pulsations \citep{Archibald2009,  Papitto2013, Tendulkar2014}, providing a link between LMXBs and spider pulsars. Therefore, it is possible that the ablation of the companion star already begins in the accreting LMXB phase, in which case EXO 0748$-$676 may be described as a transitional redback pulsar\footnote{\citet{Ratti2012} labelled EXO 0748$-$676 as \textit{black widow - like} but, as was noted by \citet{Parikh2021}, the name \textit{redback-like} would have been more appropriate given the likely companion mass.}.

Here we model X-ray eclipse profiles of EXO 0748$-$676 in multiple energy bands, using archival \textit{XMM-Newton} data. The ingress and egress durations are influenced by the size of the X-ray source and any atmosphere or structure surrounding the companion star. In one limiting case whereby the companion star is an optically thick sphere with a sharp boundary, the eclipse profiles constrain the energy-dependent radius of the X-ray emitting region and therefore can be used to place limits on the NS radius. In the opposite limiting case of an X-ray point source, the eclipse profiles instead probe the structure of the companion star's surroundings. We find that the energy-dependent eclipse profiles require a narrow, asymmetric layer of absorbing material to surround the companion star, consistent with the transitional redback pulsar scenario. Absorption and scattering in this material layer dominate the ingress and egress duration meaning that we cannot constrain the NS radius. Our model does, however, constrain $q$, and therefore $M_{\rm ns}$. In Section \ref{Section:Data}, we detail our data reduction procedure before presenting stacked energy-resolved eclipse profiles and a fit to the time-averaged spectrum. In Section \ref{Section:TRS}, we model the orbital phase-resolved spectra of the eclipse ingress and egress. In Section \ref{Section:ECProfs}, we model the energy-resolved eclipse profiles and use our results to derive a posterior probability distribution for the NS mass. We discuss our results in Section \ref{Section:Discussion} and conclude in Section \ref{Section:Conclusion}.

%%%%%%%%%%%%%%%%%%%%%%%%%%%%%%%%%%%%%%%%%%%%%%%%%%%%%%%%%%%%%%%%%%%%%%%%%%%%%%%%%%%%%%%%%%%
% DATA REDUCTION AND ANALYSIS
%%%%%%%%%%%%%%%%%%%%%%%%%%%%%%%%%%%%%%%%%%%%%%%%%%%%%%%%%%%%%%%%%%%%%%%%%%%%%%%%%%%%%%%%%%%
\section{Data Reduction and Analysis}
\label{Section:Data}
We consider the archival \textit{XMM-Newton} observation of EXO 0748$-$676 taken in April 2005 (Obs-ID 0212480501) when the source was in the soft spectral state \citep{Ponti2014}. The soft X-ray coverage offered by \textit{XMM-Newton} enables us to test models of absorption in any structure surrounding the companion star. During the observation, the EPIC-pn (European Photon Imaging Camera) was in timing mode and captured four full X-ray eclipses (with a total exposure of $42.48$ ks). Here we describe the data reduction procedure followed and our initial spectral and timing analysis.

\subsection{Data Reduction}
We used the \textit{XMM–Newton} Science Analysis Software (SAS) version 18.0 to reduce data from the EPIC-pn in timing mode. We generated calibrated and concatenated event lists using \textsc{epproc} with the default settings for timing mode as of SAS v18.0 (\texttt{runepreject=yes withxrlcorrection=yes runepfast=no withrdpha=yes}). We filtered the event list for flaring particle background using the SAS routine \textsc{espfilt}, and barycentered using \textsc{barycen}. The ingress of the first eclipse is heavily interrupted by the flaring particle background and is therefore filtered out. Two egresses are also impacted by this flaring, although this does not result in the total loss of either egress. We applied further standard filters to ignore bad pixels (\texttt{FLAG==0}), housekeeping events (\texttt{\#XMMEA\_EP}) and keep only single and double events (\texttt{PATTERN $\leq$ 4}). For all products, we used a source region of $31 \leq$ \texttt{RAWX} $\leq 45$, all \texttt{RAWY}; and a background region of $3 \leq$ \texttt{RAWX} $\leq 5$, all \texttt{RAWY}. We extracted spectra and light curves using \textsc{evselect} and generated response and ancillary files using \textsc{rmfgen} and \textsc{arfgen}. We re-binned all spectra to have at least 25 counts per channel using \textsc{specgroup}. We find that the source contributes 99.4 per cent of the total counts. We extracted light curves with 1 second time binning for a range of different energy bands: $0.2 - 10.0$ keV, $0.4 - 1.0$ keV, $1.0 - 2.0$ keV, $2.0 - 4.0$ keV, $4.0 - 6.0$ keV and  $6.0 - 8.0$ keV. Since the source dominates over the background, we do not perform a background subtraction for the light curves. We consider the calibration accuracy of the instrument in timing mode, concluding that our time-domain analysis in broad energy ranges will be robust to any effects. 

%%%%%%%%%%%%%%%%%%%%%%%%%%%%%%%%%%%%%%%%%%%%%%%%%%%%%%%%%%%%%%%%%%%%%%%%%%%%%%%%%%%%%%%%%%%
% LIGHT CURVES AND T90/T10
%%%%%%%%%%%%%%%%%%%%%%%%%%%%%%%%%%%%%%%%%%%%%%%%%%%%%%%%%%%%%%%%%%%%%%%%%%%%%%%%%%%%%%%%%%%
\subsection{Eclipse profiles}
\label{sec:profiles}
\begin{figure}
\centering
\includegraphics[width=\columnwidth]{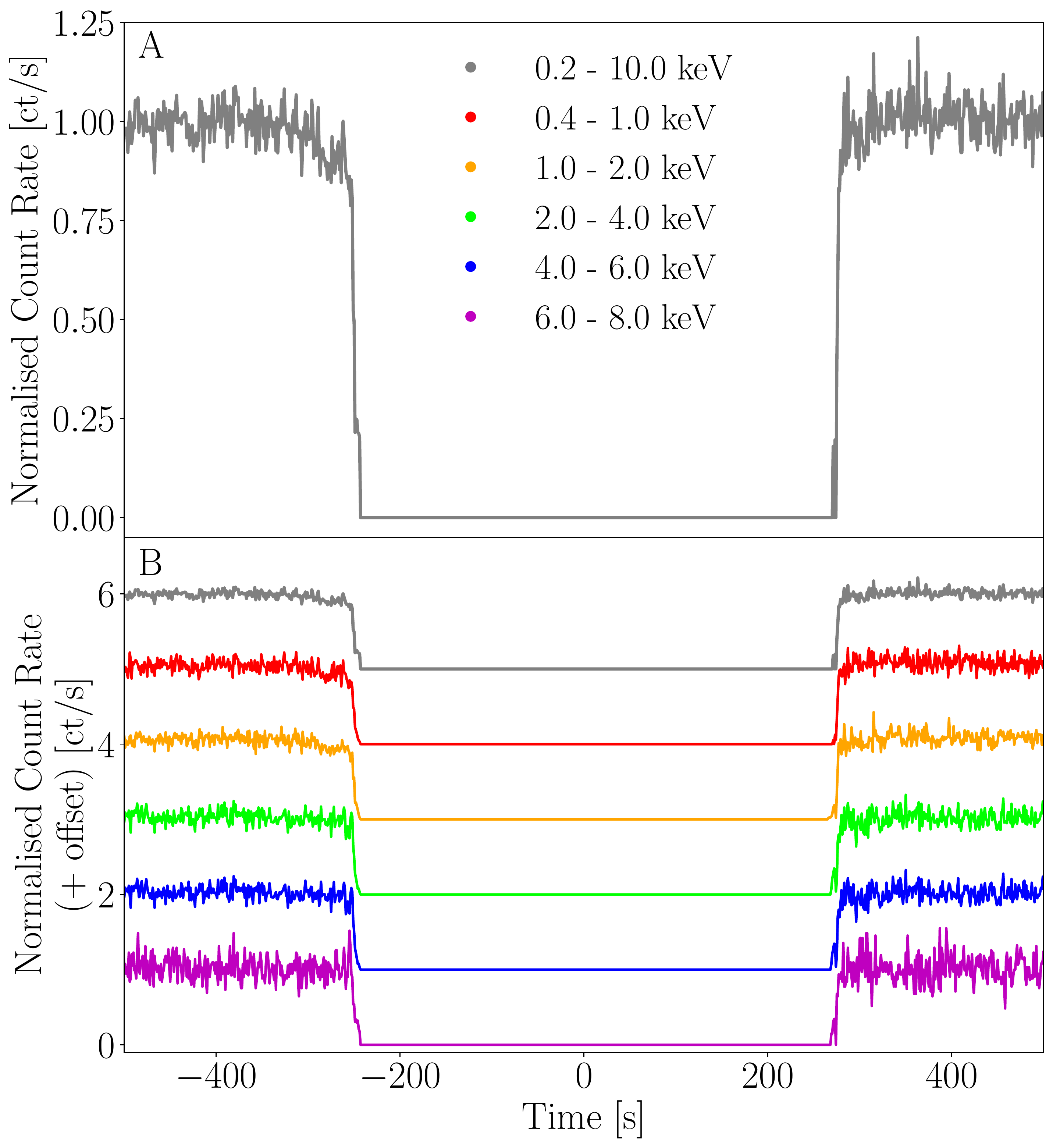}
\vspace{-0.5cm}
\caption{The folded eclipse profile of EXO 0748$-$676 in the soft-state, seen for the full energy range of {\it XMM-Netwon} ($0.2-10.0$ keV, grey) in Panels A and B, and for narrower energy ranges $0.4 - 1.0$ keV (red), $1.0 - 2.0$ keV (orange), $2.0 - 4.0$ keV (green), $4.0 - 6.0$ keV (blue) and $6.0 - 8.0$ keV (magenta) in Panel B. For all eclipse profiles the count rate is normalised by dividing through by the mean out-of-eclipse count rate such that the out-of-eclipse level is $1.0$ and the totality level is $0.0$. In Panel B, eclipse profiles are displayed with a vertical offset for visual clarity. These are $+0.0$ (magenta), $+1.0$ (blue), $+2.0$ (green), $+3.0$ (orange), $+4.0$ (red) and $+5.0$ (grey).}
\label{fig:ECProfs}
\end{figure}

\begin{figure*}
\centering
\includegraphics[width=0.95\textwidth]{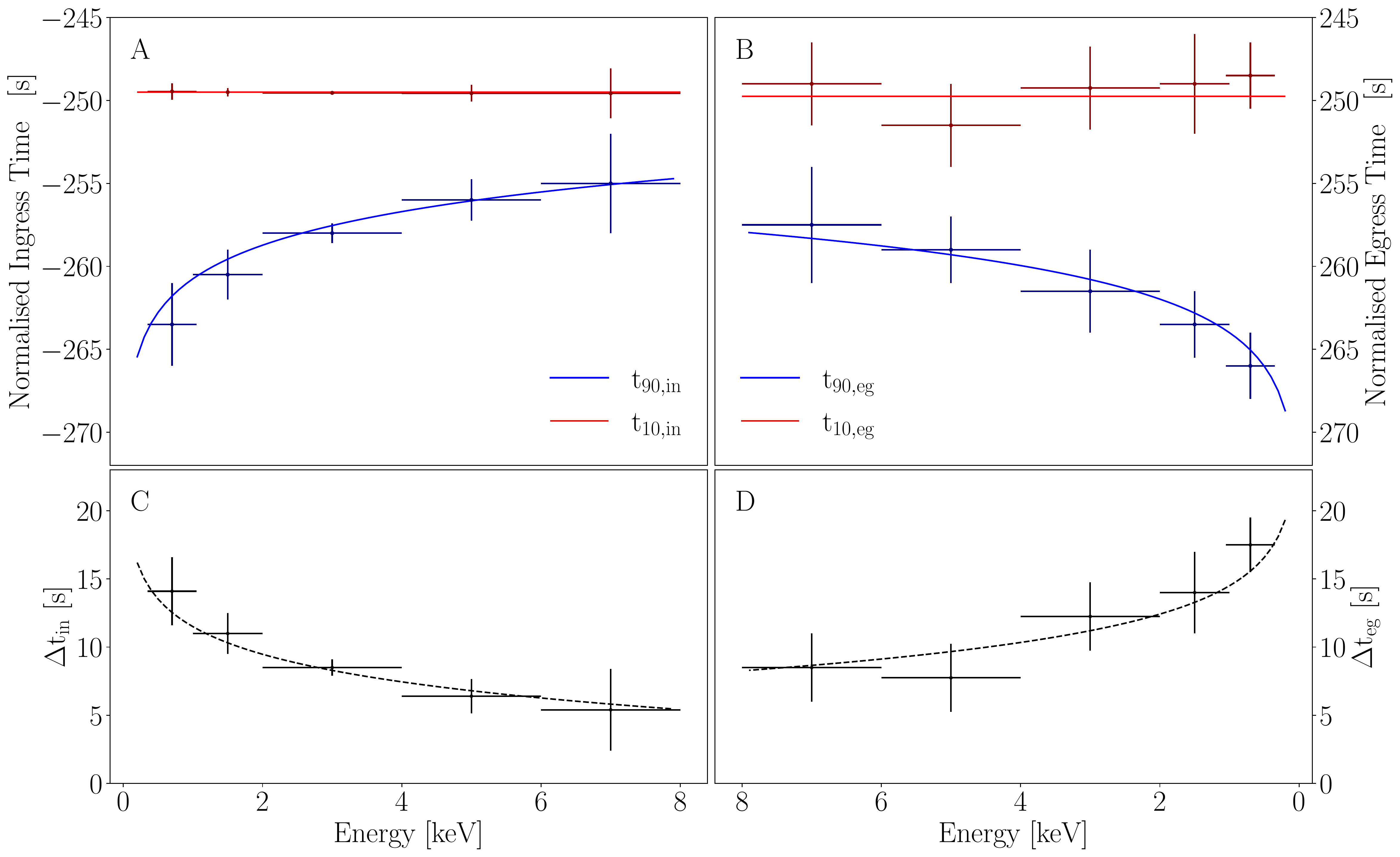}
%\vspace{-0.5cm}
\caption{Measured eclipse times, $\rm{t}_{90}$ and $\rm{t}_{10}$, as functions of energy for the ingress and egress; panels A and B respectively. We define the times $\rm{t}_{90}$ and $\rm{t}_{10}$ as, respectively, the time at which the count rate is first at 90 and 10 per cent of its mean out-of-eclipse level. Times are measured from each of the eclipse profiles in Figure \ref{fig:ECProfs} B. Panel A: Ingress start times ($\rm{t}_{90, \rm{in}}$) increase with energy indicating that eclipses start later for higher photon energies. Ingress end times ($\rm{t}_{10, \rm{in}}$), which mark the start of totality are independent of energy. Panel B: The end of totality that marks the start of the egress ($\rm{t}_{10, \rm{eg}}$) is approximately independent of energy, but the egress ends later for softer X-rays ($\rm{t}_{90, \rm{eg}}$). Note that both axes have been reversed to aid the comparison of the egress with the ingress. Panels C and D respectively show that the ingress and egress duration increases with photon energy.}
% the energy dependent ingress and egress duration. 
% The ingress starts later for higher energies but ends at the same time for all energies, therefore the ingress duration decreases for increasing energy. The egress end later for lower energies but begins at roughly the same time for all energies, therefore the egress duration increases for decreasing energy.
\label{fig:t90t10}
\end{figure*}

\begin{figure}
\centering
\includegraphics[width=\columnwidth]{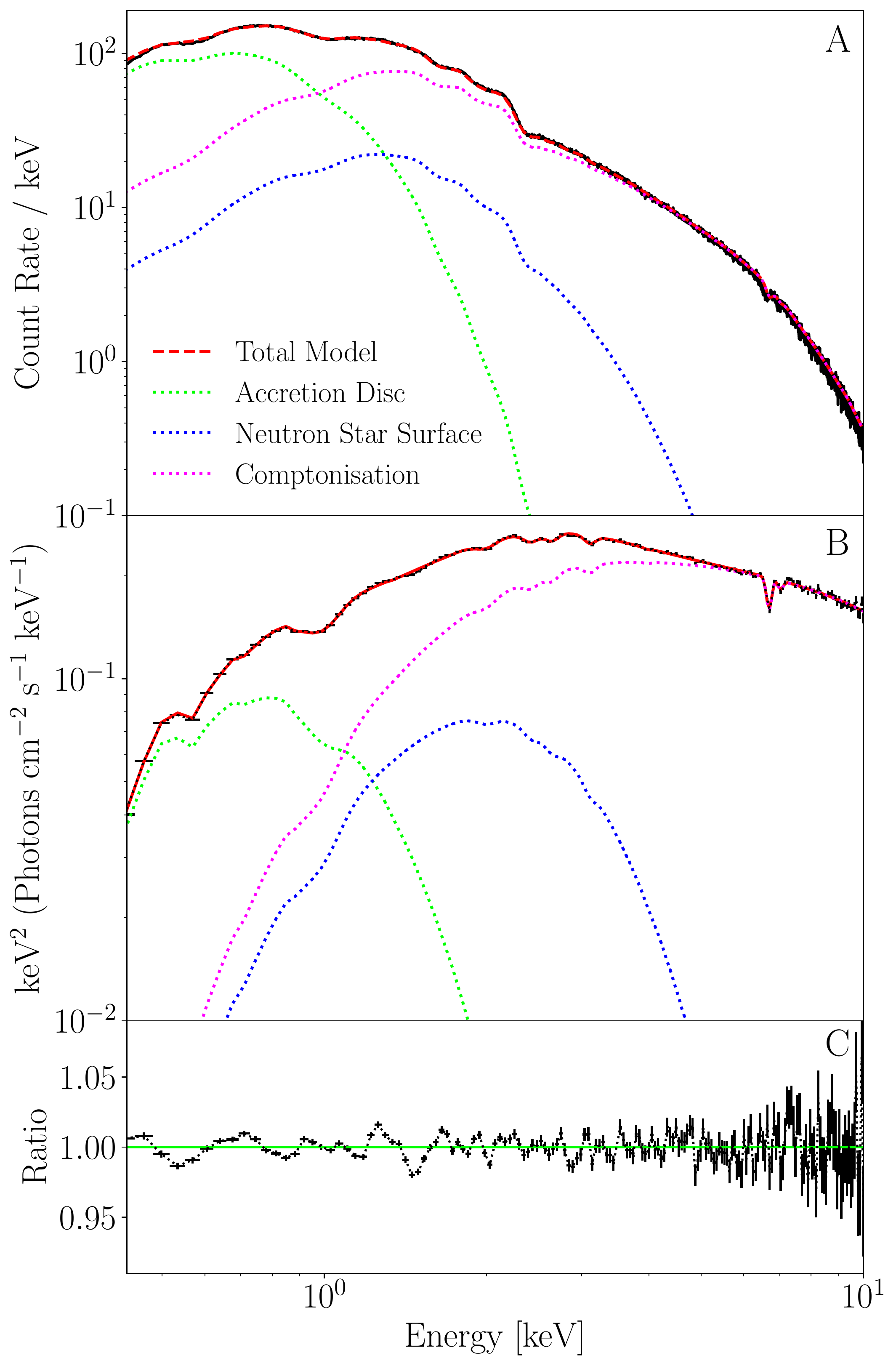}
\vspace{-0.5cm}
\caption{A fit to the time-averaged spectrum of EXO 0748$-$676 (black) using the multi-component model (red) described in Section \ref{Section:SpecFit}. Also shown are the model components originating from the NS's surface (blue), the accretion disk (green) and the thermal Comptonisation component (magenta). Model parameters are summarised in table \ref{tb:specpars} % $\chi^{2}_{\nu} = 1.07$. 
Panel A shows the best-fitting folded spectrum, panel B shows the best-fitting unfolded spectrum and panel C shows the ratio: data/folded model.}
\label{fig:Spectrum}
\end{figure}

We fold the extracted light curves on the orbital period of 3.824 hours and divided through by the mean out-of-eclipse count rate. Figure \ref{fig:ECProfs} shows the resulting eclipse profiles that are normalised to have an out-of-eclipse count rate of $1.0$ and a totality level of $0.0$. The full $0.2 - 10.0$ keV band eclipse profile (Figure \ref{fig:ECProfs}A) shows an initially gradual decline towards totality between the normalised count rate $0.8 - 1.0$ ct/s. This behaviour is mirrored in the egress, which shows a gradual rise out of totality between the normalised count rate $0.8 - 1.0$ ct/s. Figure \ref{fig:ECProfs}B shows the eclipse profile for five energy bands, with the full band profile reproduced for comparison. To investigate the energy dependence of the eclipse profiles, we define the times $\rm{t}_{90}$ and $\rm{t}_{10}$ as those at which the count rate is first at 90 and 10 per cent of the mean out-of-eclipse level, respectively. Therefore, for the ingress, $\rm{t}_{90}$ marks the start and $\rm{t}_{10}$ the end, whereas for the egress, $\rm{t}_{10}$ marks the start and $\rm{t}_{90}$ the end. To account for stochastic variability, we define these times as the time when the average count rate first passes the desired percentage and stays past it for at least five seconds. The measured $\rm{t}_{90}$ and $\rm{t}_{10}$ values for the ingress and egress are shown in Figure \ref{fig:t90t10}, panels A and B respectively. For the ingress, we see that $\rm{t}_{90}$ increases with energy whereas $\rm{t}_{10}$ is approximately constant, indicating that the ingress begins later for higher photon energies, but totality starts at approximately the same time for all energies. This behaviour is mirrored in the egress: the end of totality is approximately independent of energy, but the egress ends later for softer X-rays. The ingress and egress durations, therefore, decrease with increasing photon energy. This is shown explicitly in Figures \ref{fig:t90t10}C and \ref{fig:t90t10}D. We see from these plots that the egress is longer in duration than the ingress. In the full band, the ingress duration is approximately $\rm{t_{10, in, 0.2-10.0 keV} - t_{90, in, 0.2-10.0 keV}} = 15.2$ s and the egress duration is approximately $\rm{t_{90, eg, 0.2-10.0 keV} - t_{10, eg 0.2-10.0 keV}} = 17.5$ s. 
%These times are 14.3 s and 16.0 s respectively when considering the energy range $0.4 - 8.0$ keV, as we do in the majority of our analysis.   

% The ingress and egress duration's are also found to depend on energy. Since the ingress ends at roughly the same time for all energies but starts later for higher energies, the ingress duration decreases for increasing energies. The same is true for the egress which shows longer egress duration's for lower energies and shorter duration's for higher energies. We show the energy dependent ingress and egress duration's in Figures \ref{fig:t90t10}C and Figure \ref{fig:t90t10}D respectively. 
% with NS mass $M_{\rm ns}=1.4~M_\odot$ and
% $r_{\rm x} \approx 3700~{\rm km}~\approx 1800~r_{\rm g}$,

\subsection{Interpretation}
The eclipses result from the companion star passing in front of the X-ray emitting region close to the NS. To explain the extended ingress and egress duration, we consider two limiting cases: 1) an extended X-ray emitting region eclipsed by an optically thick companion star with a sharp outer boundary. 2) a point-like X-ray source eclipsed by a companion star surrounded by a layer of absorbing material with some radial density profile. 

\subsubsection{Extended Source with an Optically Thick Companion}
In this case, the ingress duration is the time it takes for the companion to move across our view of an extended X-ray emitting region. For an edge-on, circular binary system, we can relate the radius of the assumed spherical X-ray emitting region, $r_{\rm x}$, to the ingress duration, $\Delta t_{\rm in}$, as $r_{\rm{x}} = \pi r_{\rm a } \Delta t_{\rm in} / P$, where $P$ is the orbital period and $r_{\rm a}$ is the binary separatiosn. Calculating $r_{\rm a}$ from Kepler's law and assuming a mass ratio of $q=0.2$, we find that the observed ingress duration of $\Delta t_{\rm in}=15.2$ s requires an X-ray emitting region radius of $r_{\rm x} \approx 3500~{\rm km}~\approx 1700~r_{\rm g}$ for $M_{\rm ns} = 1.4~M_\odot$ and $r_{\rm x} \approx 4200~{\rm km}~\approx 1200~r_{\rm g}$ for $M_{\rm ns} = 2.4~M_\odot$; where $r_{\rm g}=GM_{\rm ns}/c^2$ is a gravitational radius. 

In Appendix \ref{Section:rx}, we show that $r_x$ and $q$ can be inferred as a function of binary inclination angle from the ingress duration and totality duration, under the assumption that the companion is filling its Roche-Lobe (Figure \ref{fig:rxrg}). The minimum inclination for which there is a solution is $i \approx 69^\circ$, corresponding to a mass ratio of $q=1$. For a more realistic mass ratio of $q\lesssim 0.4$, we find $i \gtrsim 73^\circ$. From the minimum possible inclination, we find a minimum X-ray emitting region size of $r_x \approx 1100~{\rm km} \approx 550~r_g$ for $M_{\rm ns} = 1.4~M_\odot$ and $r_x \approx 1400~{\rm km} \approx 400~r_g$ for $M_{\rm ns} = 2.4~M_\odot$. This large minimum source size is incompatible with, e.g., the high blackbody temperature that we measure ($kT \approx 0.5$ keV, $T \approx 5.7 \times 10^{6}$ K; see Table \ref{tb:specpars}), which would require a luminosity of $\sim 50$ times the Eddington limit, and for the system to be located outside of the Galaxy.

We observe the ingress and egress duration to decrease with increasing photon energy. This can be explained by the extended source model if the inner region of the source emits a harder spectrum than the outer region. In this case, the companion star starts to block the soft X-ray emitting region before it starts to cover the hard X-ray emitting region such that the hard X-ray ingress would begin after the soft X-ray ingress. However, this simple scenario predicts the hard X-ray ingress would end \textit{before} the soft X-ray ingress. In contrast, we observe the start and end of totality to be roughly independent of energy, therefore this scenario cannot reproduce the energy dependence of the eclipse profiles. It also cannot reproduce the egress being longer in duration than the ingress, as we observe it to be.

\subsubsection{Point Source with a Material Layer Surrounding the Companion}
In the opposite limiting case, the ingress and egress occur when our view of a point-like X-ray source is blocked by a layer of absorbing material surrounding the companion star. In this picture, the surrounding material absorbs soft X-rays more efficiently than hard X-rays at the start of the ingress until the column density becomes very large at the end of the ingress, therefore, reproducing the observed energy dependence of the ingress and egress. If the absorbing material trails somewhat behind the companion star, this can explain why we observe the egress to be consistently longer than the ingress (as seen for many more eclipses: \citealt{Wolff2009, Parmar1991}). This is the model we adopt in this paper. We assume an X-ray point source throughout, because in this model a point source is indistinguishable from an extended source unless the source is hundreds of km across or larger. 

\vspace{-0.1cm}
\subsection{Fit to the Time-Averaged Spectrum}
\label{Section:SpecFit}

\begin{table}
\begin{center}
\begin{tabular}{ c  c  c  c } 
\hline
Model Component & Parameter & Value & 1$\sigma$ \ Interval \\
\hline
\texttt{TBabs} & $\rm{N}_H$ [1 $\times$ 10$^{22}$ cm$^{-2}$] & 0.149 & $\pm^{0.006}_{0.006}$ \\ 
\hline
\texttt{diskbb} & T$_{\text{in}}$ [keV] & 0.218 & $\pm^{0.006}_{0.006}$\\ 
\hline
\texttt{bbody} & $kT$ [keV] & 0.491 & $\pm^{0.010}_{0.012}$ \\ 
\hline
\vspace*{0.1cm}
\texttt{NthComp} & $\Gamma$ & 2.091 & $\pm^{0.026}_{0.053}$\\
& $kT_e$ [keV] & 3.469 & $\pm^{0.418}_{0.310}$ \\ 
\hline
\vspace*{0.1cm}
Absorption lines & $E_1$ [keV] & 6.687 & $\pm^{0.009}_{0.008}$ \\ 
\vspace*{0.1cm}
& $E_2$ [keV] & 7.021 & $\pm^{0.030}_{0.021}$ \\ 
\vspace*{0.1cm}
& $E_3$ [keV] & 2.639 & $\pm^{0.021}_{0.020}$ \\ 
\vspace*{0.1cm}
& $E_4$ [keV] & 3.111 & $\pm^{0.018}_{0.018}$ \\ 
\vspace*{0.1cm}
& $E_5$ [keV] & 3.979 & $\pm^{0.229}_{0.127}$ \\ 
\vspace*{0.1cm}
 & $\sigma_{1-5} [keV] $ & 0.007 & $\pm^{0.005}_{0.005}$ \\
\hline
\vspace*{0.1cm}
Calibration lines  & $E_6$ [keV] & 0.990 & $\pm^{0.004}_{0.004}$ \\ 
\vspace*{0.1cm}
 & $\sigma_{6} [keV] $ & 0.070 & $\pm^{0.002}_{0.002}$ \\
\vspace*{0.1cm}
& $E_7$ [keV] & 2.002 & $\pm^{0.016}_{0.016}$ \\ 
\vspace*{0.1cm}
 & $\sigma_{7}$ [keV] & 0.002 & $\pm^{0.001}_{0.001}$ \\
\vspace*{0.1cm}
& $E_8$ [keV] & 2.419 & $\pm^{0.018}_{0.019}$\\
\vspace*{0.1cm}
 & $\sigma_{8}$ [keV] & 0.016 & $\pm^{0.002}_{0.001}$ \\
\hline
\end{tabular}
\caption{\label{tb:specpars}Best-fitting parameters from our fit to the time-averaged spectrum. Line energies quoted are centroid energies. Following \citep{Ponti2014}, the first 5 lines correspond to 1) a Fe XXIII-XXV K$\alpha$ blend, 2) Fe XXVI K$\alpha$, 3) S XVI K$\alpha$, 4) a S XVI K$\beta$ and Ar XVII K$\alpha$ blend, 5) A Ca XIX-XX K$\alpha$ blend. The final 3 lines account for calibration issues (see e.g. \citealt{DeMarco2016}). Reduced $\chi^2$ is $\chi^2/\nu=161.57/151$. Since the duration of the ingress and egress are only$ \sim0.01\%$ of the orbital period each, and no counts are contributed during totality, the time-averaged spectrum is approximately the out-of-eclipse spectrum.}
\end{center}
\end{table}

We fit the time-averaged spectrum using \texttt{xspec} V12.11.1 \citep{Arnaud1996} and the model
\begin{equation}
    \mathtt{TBabs*(diskbb + bbody + NthComp) * G}.
\end{equation}
Here, \texttt{TBabs} accounts for absorption by the interstellar medium (we use the abundances of \citealt{Wilms2000}), and \texttt{diskbb} is a multi-temperature accretion disk spectrum. We model the spectrum from the NS surface as a blackbody (\texttt{bbody}) plus Comptonisation by a thermal population of electrons with temperature $kT_{\rm e}$ (\texttt{NthComp}; \citealt{Z1996}). We tie the seed photon temperature to the blackbody temperature, physically corresponding to some fraction of the NS surface blackbody photons being Compton up-scattered. Finally, \texttt{G} represents eight Gaussian absorption lines (\texttt{gabs}). The first five of these correspond to astrophysical absorption lines, originally discovered in this observation by \citet{Ponti2014}. We tie the widths of these five lines but leave their centroids and strengths as free parameters. Several calibration features are evident in the $E < 2.5$ keV region of the spectrum, as is common for the EPIC-pn (see e.g. \citealt{DeMarco2016}). These features motivated \citet{Ponti2014} to ignore energies below $2.5$ keV. However, we wish to model the eclipse profile in soft X-rays, for which we need to extend our model for the out-of-eclipse spectrum to lower energies. We, therefore, account for the calibration features with three additional \texttt{gabs} components. 

Figure \ref{fig:Spectrum} shows the total model (red) and its constituent components: the blackbody from the NS's surface (blue), the multi-temperature blackbody from the accretion disk (green) and the thermal Comptonisation component (magenta). We achieve an acceptable fit with a reduced $\chi^2$ value of $\chi^2/\nu = 161.57/151$ and a null hypothesis probability, $p = 0.426$. The best-fitting parameters are reported in Table \ref{tb:specpars}. Since the duration of the ingress and egress are only $\sim 0.01 \%$ of the orbital period each, and no counts are contributed during totality, the time-averaged spectrum is approximately the out-of-eclipse spectrum and is subsequently used in our eclipse profile modelling.

\section{Phase-Resolved Spectroscopy}
\label{Section:TRS}

\begin{figure*}
\centering
\includegraphics[width=0.95\textwidth]{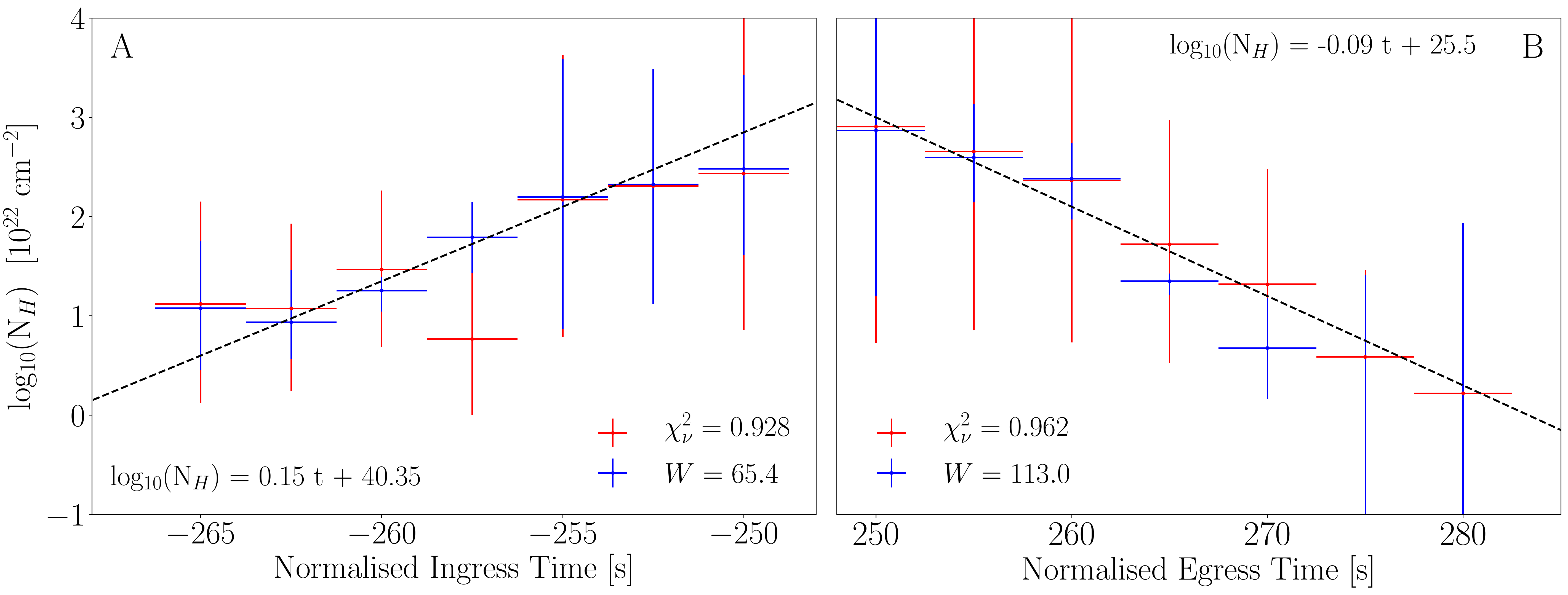}
%\vspace{-0.5cm}
\caption{Hydrogen column densities as functions of time for the ingress (Panel A) and egress (Panel B), obtained by fitting our local absorption and scattering model \texttt{abssca} to 7 phase-resolved ingress spectra and 7 phase-resolved egress spectra. We present results obtained using two fit statistics - $\chi^{2}$ (red), quantified by $\chi^{2} / \nu$ and C-Stat (blue), quantified by the $W$ statistic. As shown, the measured Hydrogen column densities are found to be independent of the chosen fit statistic but note that C-Stat is more appropriate here because of the low count rate in several of the spectra. The eclipse times are normalised such that the centre of the eclipse occurs at $t=0$ s. In both A and B, the black dashed line shows the log-linear trend of the data. The corresponding equations for each line are given in their respective panels.}
\label{fig:TRS}
\end{figure*}

\begin{table}
\begin{center}
\begin{tabular}{ c  c  c  c  c } 
\hline
Parameter & Ingress & & Egress &\\
& $\frac{\chi^{2}}{\nu}$ = $\frac{40.8}{44}$ & $W$ = 65.4 & $\frac{\chi^{2}}{\nu}$ = $\frac{42.3}{44}$ & $W$ = 113.0 \\ 
& $p = 0.77$ & $p=0.67$  & $p = 0.54$ & $p = 0.47$ \\
\hline
\hline
$\rm{f_{cov}}$ & 0.99 \ $\pm^{0.01}_{0.01}$ & 1.00 \ $\pm^{0.00}_{0.01}$ & 0.99 \ $\pm^{6 \times 10^{-3}}_{0.01}$  & 1.00 \ $\pm^{0.00}_{0.01}$ \\ 
\hline
$\rm{log(\xi)}$ & 2.81 \ $\pm^{0.15}_{0.55}$ & 3.21 \ $\pm^{0.08}_{0.08}$ & 2.57 \ $\pm^{0.83}_{0.19}$ & 2.80 \ $\pm^{0.42}_{0.65}$ \\ 
\hline
\vspace*{0.1cm}
${\rm{N}_{H}}$ & 13.2 \ $\pm^{10.8}_{9.93}$ & 12.0 \ $\pm^{0.24}_{0.21}$ & 808 \ $\pm^{80.8}_{151}$ & 739 \ $\pm^{42.6}_{46.7}$  \\ 
\vspace*{0.1cm}
$[10^{22}\rm{cm}^{-{2}}]$ & 11.9 \ $\pm^{7.16}_{6.84}$ & 8.62 \ $\pm^{0.42}_{0.30}$ & 453 \ $\pm^{145}_{63.4}$ & 394 \ $\pm^{3.45}_{2.83}$\\ 
\vspace*{0.1cm}
& 29.4 \ $\pm^{6.25}_{6.00}$ & 18.0 \ $\pm^{1.37}_{1.63}$ & 231 \ $\pm^{41.7}_{42.8}$ & 242 \ $\pm^{2.29}_{2.57}$ \\ 
\vspace*{0.1cm}
& 5.85 \ $\pm^{6.35}_{5.85}$ & 62.1 \ $\pm^{2.25}_{2.27}$ & 52.8 \ $\pm^{17.8}_{15.8}$ & 22.4 \ $\pm^{1.20}_{1.38}$\\ 
\vspace*{0.1cm}
& 148 \ $\pm^{28.6}_{24.1}$ & 158 \ $\pm^{24.6}_{21.6}$ & 20.8 \ $\pm^{14.4}_{12.9}$ & 4.75 \ $\pm^{0.30}_{0.32}$ \\ 
\vspace*{0.1cm}
& 203 \ $\pm^{10.6}_{10.4}$ & 212 \ $\pm^{14.7}_{16.0}$ & 3.86 \ $\pm^{7.58}_{3.86}$ & $10^{-9}$ \ $\pm {10^{-10}}$\\ 
\vspace*{0.1cm}
& 272 \ $\pm^{54.7}_{37.8}$ & 303 \ $\pm^{8.87}_{7.37}$ & 1.66 \ $\pm^{10.6}_{1.66}$ &  $10^{-8}$ \ $\pm {10^{-10}}$ \\ 
\hline
\hline
$n_{\rm{H}}$ & & 0.56 $\pm^{0.08}_{0.07}$ & & 0.56 $\pm^{0.08}_{0.07}$ \\
$[10^{22}\rm{cm}^{-{2}}]$ & & &  \\
\hline
$\alpha$ & & 2.68 $\pm^{0.14}_{0.15}$ & & 2.68 $\pm^{0.14}_{0.15}$\\
\hline
\end{tabular}
\caption{\label{tb:TRSPars} Best fitting parameters obtained from fitting our local absorption and scattering model \texttt{abbsca} to the phase-resolved spectra of the ingress and egress. We present results obtained using two fit statistics - $\chi^{2}$, quantified by $\chi^{2} / \nu$ and C-Stat, quantified by the $W$ statistic. Their values are given in the first and second columns respectively for both the ingress and egress, with corresponding null hypothesis probabilities below. The model parameters are found to be consistent between the two fit statistics, but C-Stat is more appropriate because of the low count rate in several of the spectra. Here, $\chi^{2}_{\nu}$ provides a simple way to understand the goodness of fit. The rows are covering fraction, $\rm{f_{cov}}$, ionisation parameter, $\rm{log(\xi)}$, and column density for each time bin, $\rm{N_H}$. The column densities are listed chronologically. The bottom two rows detail the best-fitting parameters for the background spectral model, \texttt{TBabs*po}, which are the equivalent hydrogen column, $n_{\rm{H}}$ and the power-law index, $\alpha$. All errors are $1 \sigma$ and the same background spectral fit was applied to both the ingress and egress.}
\end{center}
\end{table}

To diagnose the nature of the absorbing material around the companion star, we fit the phase-resolved spectra of the ingress and egress with a model that accounts for the absorption and scattering of X-ray photons by the material.

\subsection{Spectral Model}
We define a local \textsc{xspec} model called \textsc{abssca} in which the total transmitted specific intensity $I_{\rm E}$ relates to the out-of-eclipse specific intensity $I^0_{\rm E}$ as
\begin{equation}
    I_{\rm E} = I^0_{\rm E} \bigg\{ \rm{f}_{\rm cov} \exp\{-N_{\rm H} [\sigma(\textit{E}) + (n_{\rm e}/n_{\rm H})\sigma_{\rm T} ] \} + 1 - \rm{f}_{\rm cov} \bigg\},
\end{equation}
where $\rm{f}_{\rm cov}$ is the covering fraction of the absorbing material around the companion star (a fraction $\rm{f}_{\rm cov}$ of the incident photons have interactions and the rest pass straight through), $\rm{N}_H$ is the hydrogen column density, $\sigma(E)$ is the absorption cross-section, $\sigma_{\rm T}$ is the Thomson electron scattering cross-section, and the density ratio $\rm{n_e/n_H}$ is the ratio of free electrons to hydrogen nuclei in the medium (i.e. the electron column density is $\rm{N}_e = \rm{N}_H~n_e / n_H$). We use the \textsc{xspec} model \textsc{zxipcf} \citep{Miller2006} to calculate the absorption cross-section, which depends on the ionisation parameter $\xi$. 

%\textsc{zxipcf} assumes that the absorbing material is ionised by incident radiation with a spectrum given by a power law with photon index $\Gamma=2$, which is approximately consistent with the out-of-eclipse spectrum.\textcolor{red}{MM: we should add something a bit more here as I wouldn't immediately conclude this from Fig 3}\cmtai{Let's just get rid of this sentence rather than opening Pandora's box.}

% , $\sigma(E) \sim \propto E^{-3}$,

In addition to being absorbed by the material surrounding the companion, photons can be scattered by the material layer out of the line of sight. Also, photons can be scattered \textit{into} the line of sight, but this effect is negligible in our case because the stellar absorber subtends a small solid angle according to the X-ray source. This is in contrast to a commonly considered scenario of a spherical shell of absorbing material surrounding a central source, in which case photons scattered into the line of sight will exactly cancel those scattered out of the line of sight. For $\xi \lesssim 100~{\rm erg}~{\rm cm}~{\rm s}^{-1}$, the absorption cross-section dominates over the scattering cross-section except for $E \gtrsim 10$ keV. For $\xi \gtrsim 100~{\rm erg}~{\rm cm}~{\rm s}^{-1}$, $\sigma(E)$ additionally dips below $\sigma_{\rm T}$ for $E \lesssim 0.5$ keV and $E \sim 7$ keV (see e.g. Figure 1.21 of \citealt{Done2010}). The density ratio $\rm{n_e/n_H}$ can, in principle, be calculated from the ionisation state and relative elemental abundances of the gas, such that $\rm{n_e/n_H}$ increases with $\xi$. For instance, a pure hydrogen gas would have $\rm{n_e/n_H}$ in the range zero to unity, whereas the presence of heavier elements makes $\rm{n_e/n_H} > 1$ possible. Here we simply fix $\rm{n_e/n_H}=1$ throughout. This is appropriate for larger values of $\xi$, but not for smaller values. However, for smaller values of $\xi$, $\sigma(E) \gg \sigma_{\rm T}$ for the entire \textit{XMM-Newton} band pass and therefore scattering is negligible, rendering the model insensitive to $\rm{n_e/n_H}$. We additionally fix the redshift to $z=0$.

\subsection{Results}

We extract phase-resolved spectra for the ingress and egress from the folded eclipse profiles in 6 energy bands: $0.2 - 0.5$ keV \footnote{We chose to include this energy band in the phase-resolved spectral analysis because soft X-ray photons appear to be most susceptible to absorption by the surrounding material. For the eclipse profile modelling, it is excluded due to high variability and a low count rate.}, $0.5 - 1.0$ keV, $1.0 - 2.0$ keV, $2.0 - 4.0$ keV, $4.0 - 6.0$ keV and $6.0 - 8.0$ keV (which were calculated following the procedure described in Section \ref{sec:profiles}). Spectra are extracted using 2.5 second and 5.0 second time bins, respectively, for the ingress and egress, covering time ranges of t $=-267.5$s to t $=-250.0$s and t $=247.5$s to t $=282.5$s, producing a total of 14 spectra (seven for the ingress and seven for the egress). The time ranges are defined to ensure we fully encapsulate the times in which the absorbing material is influencing the eclipses, and to account for the observed asymmetry in the eclipse profiles. 

Because of the asymmetry in the eclipses, we allow the ingress and egress to have different values of covering fraction ($\rm{f}_{\rm cov}$), ionisation parameter ($\xi$) and column density ($\rm{N}_{H}$). We initially allow the covering fractions and ionisation parameters to be free for each spectrum, finding their values to be approximately constant across the sets of ingress and egress spectra. We, therefore, tie the covering fractions and ionisation parameters for the sets of ingress and egress spectra (such that it is constant in both sets), but the column density remains a free parameter for each spectrum. The \texttt{xspec} fit results are shown in Table \ref{tb:TRSPars}, where the column densities are listed chronologically. We find that $\log(\rm{N}_H)$ increases linearly with time during the ingress and decreases linearly with time in the egress, as shown by the dashed black lines in Figure \ref{fig:TRS}. The column density is found to change by several orders of magnitude over a short time, suggesting the absorbing material possesses a steep density profile. 

Note that two fit statistics are considered here. Due to the low count rate in several of the spectra, C-Stat \footnote{C-Stat is a likelihood-based statistic for low count-rate, Poisson distributed data.} is more appropriate than $\chi^2$. Within \texttt{xspec}, this requires a background spectrum and model to be defined because the difference between two Poisson variables is not another Poisson variable. Therefore, instead of subtracting the background from the source, the combined likelihood for the source and background are found and quantified with the \textit{W-Statistic} (see \citealt*{Arnaud1996} Appendix B for further details). The background spectrum is simply the totality spectrum and is well-modelled by an ISM absorption and power-law model: \texttt{TBabs*po}. We present best-fitting model parameters for the ingress and egress spectra using both fit statistics (C-Stat and $\chi^{2}$) in Table \ref{tb:TRSPars}, to show that the best-fitting parameters are independent of the chosen fit-statistic. Table \ref{tb:TRSPars} also details the best-fitting parameters for the background spectral fit which was used in both the ingress and egress spectral fits. %and to provide a simple way to understand the goodness-of-fit - MM: I don't see quite how it does that but feel free to add this back in if you feel strongly about it. 

\subsection{Inferring the Density Profile}
\label{sec:nx}

We model the companion star as spherically symmetric with radial hydrogen number density profile $n_\star (r)$. The photospheric radius of the star is $R_{\rm cs}$, such that $n_\star (r)$ in the region $r<R_{\rm cs}$ is large enough for the optical depth to be effectively infinite. Totality therefore occurs whenever the projected separation on the image plane between the centre of the companion star and the X-ray point source is less than $R_{\rm cs}$. Defining the \textit{impact parameter}, $b(t)$ as this projected separation in units of $R_{\rm cs}$ means that totality occurs for all times when $b(t) \leq 1$. For $b(t) > 1$, the hydrogen column density for a sight-line through the surrounding material is: 
\begin{equation}
    \rm{N}_H(t) = \int_{-\infty}^{+\infty} n_\star(r) ds,
\end{equation}
where $s$ is the distance a ray has travelled along a given sight line. Here $s=0$ is defined as when the ray passes closest to the centre of the companion star. Therefore, a ray travelling along a sight line that starts behind the star and points towards the observer extends from $s=-\infty$ to $s=+\infty$. Defining $x \equiv r / R_{\rm cs}$ as distance in units of the companion star radius, it can be shown that $(s/R_{\rm cs})^2 = x^2 - b^2$, and we can therefore re-write the above integral as
\begin{equation}
    \rm{N}_H(t) = 2 N_{H,0} \int_{b(t)}^{x_{\rm out}} n(x) \frac{x}{\sqrt{x^2-b^2(t)}} dx,
    \label{eqn:Nht}
\end{equation}
where $n(x) \equiv n_\star(r)/n_0$ such that $n_0=n_\star(r=R_{\rm cs})$ and $N_{H,0}=R_{\rm cs}n_0$ is the column density of a sight-line of length $R_{\rm cs}$ through material with constant density $n_0$. Here, $x_{\rm out}$ is the outermost radius of the absorbing medium with non-negligible density.

%\textcolor{red}{MM: this can't be correct can it as the units are then number density squared?}
%\cmtai{$n(x)$ is dimensionless. Maybe it would be clearer if we said: `where $n(x) \equiv n_\star(r)/n_0$ '} such that $n_0=n_\star(r=R_{\rm cs})$ and $N_{H0}=R_{\rm cs}n_0$ 
%\textcolor{red}{MM:can we maybe change H0 to H,0 - I don't expect anyone to think it's Hubble's constant but maybe just in case}

For a perfectly circular binary system with orbital period, $P$, separation, $r_{\rm a}$, and inclination angle to the observer, $i$, we can write the impact parameter as:
\begin{equation}
    b(\phi) = \frac{r_a}{R_{\rm cs}} \sqrt{ 1 - \sin^2 i ~\cos^2 \phi },
    \label{eqn:b}
\end{equation}
and the orbital phase as $\phi=(2\pi/P)(t-t_0)$, where $t_0$ is the time at the centre of totality. From this, we can show that the duration of totality, $t_e$, obeys
\begin{equation}
\left( \frac{R_{\rm cs}}{r_{\rm a}} \right)^2 = 1 - \cos^2\left( \frac{\pi \  t_{e}}{P} \right) ~\sin^2 i.
\label{eqn:econd}
\end{equation}
Assuming the companion star is filling its Roche Lobe, the ratio $R_{\rm cs}/r_{\rm a}$ is \citep{Horne1985, Ratti2012}:
\begin{equation}
    \frac{R_{\rm cs}}{r_a} = h(q) = \frac{0.49 q^{2/3}}{0.6 q^{2/3} + \ln(1+q^{1/3})},
    \label{eqn:RLfill}
\end{equation}
where $q=M_{\rm cs}/M_{\rm ns}$. We can then calculate the inclination angle from Equations \ref{eqn:econd} and \ref{eqn:RLfill} as:
\begin{equation}
    \sin i = \frac{\sqrt{1-h^2(q)}}{ \cos(\pi t_e / P) }.
    \label{eqn:inc}
\end{equation}
Therefore, for known $t_{\rm e}$, $P$ and $t_0$, the only model parameter required to calculate $b(t)$ is $q$. The column density $\rm{N}_H(t)$ can be calculated from an assumed radial density function, $n(x)$, with $q$ as the only other model parameter.

We trial a number of density profiles, the first is a power law:
\begin{equation}
    \centering
    n(x) = x^{-m}.
    \label{eqn:powlaw}
\end{equation}
%\textcolor{red}{MM: is there not a normalisation?}\cmtai{No, the normalisation comes in with $N_{H0}$. $n(x)$ just equals one for $x=1$.}
Setting $m=2$ corresponds to a stellar wind with constant velocity. The companion star in EXO 0748$-$676 is a low mass M-dwarf star \citep{Parmar1986} that would typically be expected to only drive a weak solar-like wind with a mass-loss rate of the order of $10^{-13}M_\odot$~yr$^{-1}$ or less \citep[e.g.,][]{Wargelin2002}. Such a wind is too low-density to yield significant absorption at X-ray wavelengths: a wind velocity of 500~km~s$^{-1}$, for example, corresponds to a particle density at the base of the wind of $6\times 10^6$~cm$^{-3}$ for the typical M-dwarf radius of $R_{cs}=0.43R_\odot$, and a column density through the wind of $2\times 10^{17}$~cm$^{-2}$. In this case, however, a much denser wind could be driven by irradiation from the NS and accretion flow.

\begin{figure*}
\centering
\includegraphics[width=0.89\textwidth]{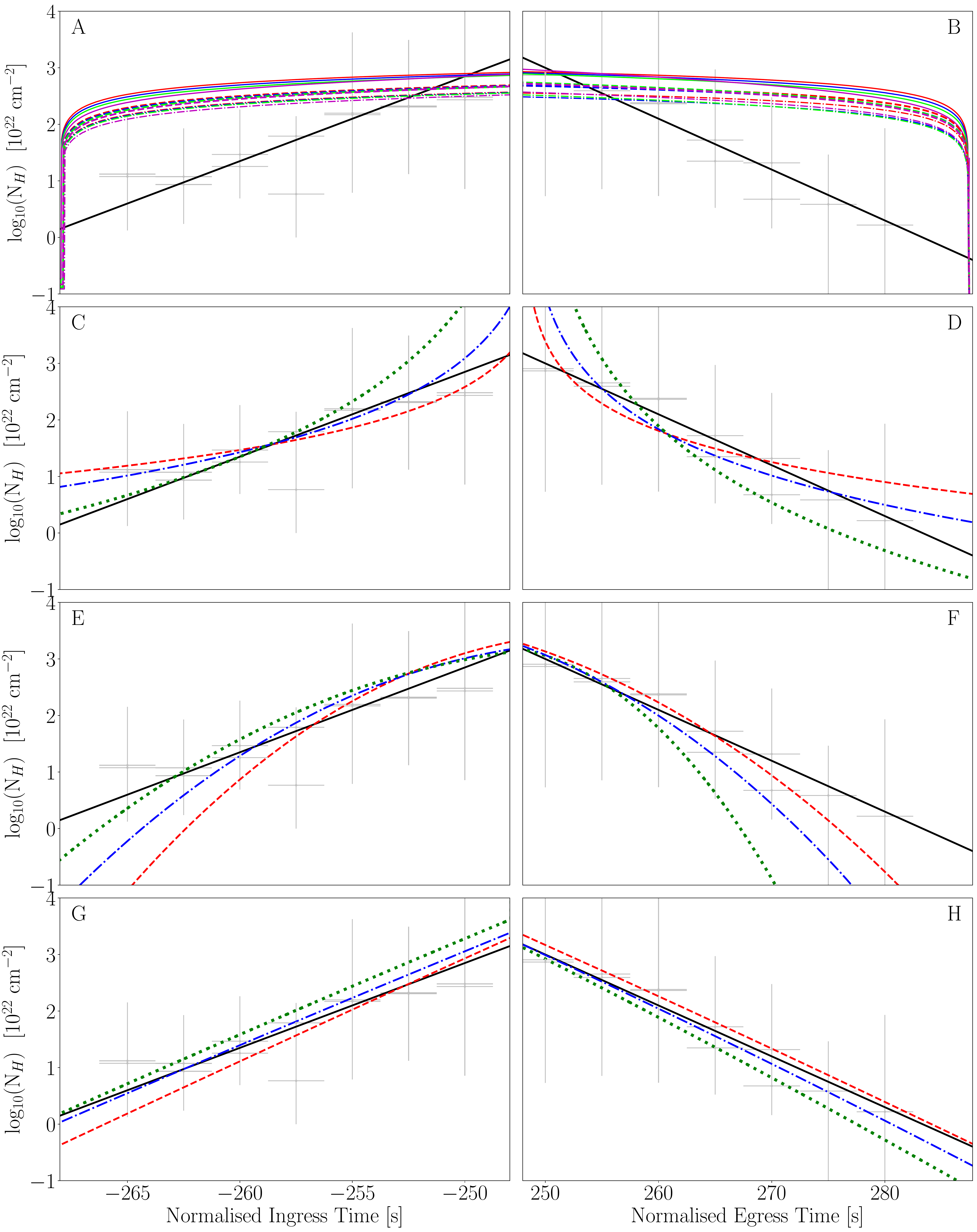}
\vspace{-0.1cm}
\caption{Panels A-H: Hydrogen column density of the absorbing material as a function of time during the ingress and egress. Measured $N_{\rm H}$ values are shown in grey with a log-linear best-fit function in black. Panels A and B assume the power law density profile (Equation \ref{eqn:powlaw}). We use mass ratio q = 0.05 (solid), q = 0.2 (dashed) and q = 0.4 (dot-dashed); which correspond to $i$ = 82.8 , $i$ = 76.9 and $i$ = 73.6 degrees respectively. We explore a power-law index of m = −2 (red), m = 0 (blue), m = 2 (green) and m = 10 (magenta). Panels C and D assume the accelerating wind profile (Equation \ref{eqn:beta}) with parameter combinations of $q = 0.05, \beta = 2.0$ (red), $q = 0.2, \beta = 3.0$(blue) and $q = 0.4, \beta = 5.0$ (green). Panels E and F assume the Gaussian density profile (Equation \ref{eqn:gauss}). We use mass ratio and fractional width parameter combinations of $q = 0.05, \Delta = 0.008$ (red), $q = 0.2, \Delta = 0.004$ (blue) and $q = 0.4, \Delta = 0.003$ (green) for the ingress and combinations of $q = 0.05, \Delta = 0.008$ (red), $q = 0.2, \Delta = 0.004$ (blue) and $q = 0.4, \Delta = 0.003$ (green) for the egress. Panels G and H assume the exponential density profile (Equation \ref{eqn:loglin}). We use mass ratio and scale height parameter combinations of $q = 0.05, h = 0.004$ (red), $q = 0.2, h = 0.002$ (blue) and $q = 0.4, h = 0.001$ (green) for the ingress and combinations of $q = 0.05, h = 0.009$ (red), $q = 0.2, h = 0.004$ (blue) and $q = 0.4, h = 0.003$ (green) for the egress. Panels C-H: We trial $q = 0.05, q = 0.2$ and $q=0.4$ for each value of $\beta$, $\Delta$ and $h$ but only plot three in each panel for visual clarity. }
\label{fig:DensProf2}
\end{figure*}

We trial power-law indices of $m = -2, \ m = 0, \ m = 2 \ \rm{and} \ m = 10$. For each index, we also explore mass ratios of $q = 0.05, \ q = 0.2 \ \rm{and} \ q = 0.4$, corresponding to inclination angles of $ i = 82.8, \ i = 76.9 \ \rm{and} \ i = 73.6$ degrees respectively. Here, $q = 0.05$ translates roughly to the canonical NS mass, $1.4 M_{\odot}$. As seen in the top row of Figure \ref{fig:DensProf2}, the power-law density profile results in an $\rm{N}_H(t)$ that overwhelmingly disagrees with observations, for both the ingress and egress. We find higher values of $m$ and $q$ to be the most consistent with $m=10$ and $q= 0.4$ providing $\chi^{2} / \nu = 464.2 / 44 $ for the ingress and $\chi^{2} / \nu = 332.4 / 44 $ for the egress. The null hypothesis probabilities for both are of the order $10^{-30}$. We conclude that the asymptotic nature of the power-law function trialled will not easily reproduce the observed, steep $N_{\rm{H}}(t)$. 
% \cmtai{I guess this only assuming a spectroscopic mass for the companion star or something? And that in turn assumes the star is main sequence? I like having this sentence in here, but I think we need to state where $M_{\rm cs}$ comes from.} \cmtak{Using the binary mass function q = 0.05 gives roughly 1.4 solar masses, so this sentence is based on that but will vary with k-correction etc. Otherwise 0.05 = $(M_{cs} \sim 0.47) / M_{ns}$ implies a 9.5 solar mass NS.  }.

The data require a density profile that drops off more steeply with radius rather than a constant velocity wind. We therefore consider a density profile corresponding to an accelerating wind \citep{Puls2008}:
\begin{equation}
    n(x) = x^{-2}~( 1 - x^{-1} )^{-\beta},
    \label{eqn:beta}
\end{equation}
where $\beta$ is a parameter that describes the acceleration. In this case, the constant $N_{\rm H,0}$ from Equation (\ref{eqn:Nht}) is related to the mass loss rate of the wind, $\dot{M}_{\rm out}$, as:
\begin{equation}
    N_{H,0} = \frac{ \dot{M}_{\rm out} }{ 4\pi \bar{m} \sqrt{G M_{\rm cs} R_{\rm cs} } },
\end{equation}
where $\bar{m}$ is the mean molecular weight of the wind material. Equation (\ref{eqn:beta}) is typically applied to massive stars for which the wind is driven by the radiation of the star itself. \citet{Puls2008} quote $\beta \leq 1.0 $ for an OB star, however, we require much higher values to obtain consistency with the data. With a parameter combination of $\beta=3.0$ and $q=0.2$, $\chi^{2} / \nu = 85.1 / 44 $, while a parameter combination of $\beta=5.0$ and $q=0.4$ gives $\chi^{2} / \nu = 64.9 / 44 $ for the ingress. For the egress these are, respectively,  $\chi^{2} / \nu = 140.1 / 44 $ and  $\chi^{2} / \nu = 74.9 / 44 $. Panels C and D of Figure \ref{fig:DensProf2} show this consistency for parameter combinations of $\beta = 2.0, \ q = 0.05, \ \beta = 3.0, \ q = 0.2, \ \rm{and} \ \beta = 5.0, \ q=0.4$, with the latter providing the best overall agreement. This could indicate that there is very fast-moving material surrounding the companion, or that this, an M-dwarf star within an X-ray binary, is an inappropriate context for this functional form (Equation \ref{eqn:beta}), especially since it is usually applied to hot, massive stars.

We consider the possibility that the absorbing material may behave less like a stellar wind and more like an extended atmosphere. Therefore we trial a Gaussian density profile characterised by the fractional width of the material layer, $\Delta$, which is a steeper function of distance, and therefore should be more consistent with the observed $\rm{N}_H(t)$. We express this form of density profile as:
\begin{equation}
    n(x) = \exp\left[ -\frac{(x-1)^2}{2\Delta^2} \right],
    \label{eqn:gauss}
\end{equation}
and try values of $\Delta = 0.008, \ \Delta = 0.005 \ \rm{and} \ \Delta = 0.003$ for the set of mass ratios detailed above. As seen in Panels E and F of Figure \ref{fig:DensProf2}, we achieve better agreement with the observed $\rm{N}_H(t)$, particularly for the ingress, where the parameter combination $\Delta = 0.003$ and $q = 0.4$ gives the most consistent result $\chi^{2} / \nu = 64.3 / 44 $. The eclipse timings (see Section \ref{sec:profiles}) suggested that there might be material trailing behind the companion star as it orbits, therefore, it is unsurprising that a larger fractional width, $\Delta = 0.008$, is most consistent with the egress data ($\chi^{2} / \nu = 64.4/ 44 $, with $q = 0.05$), further supporting the presence of an asymmetric eclipse. 
%\textcolor{red}{MM: but isn't this an even sided radial function?} \cmtak{We define $\Delta$ to be different for the ingress and egress which effectively changes the amount of atmospheric material a sight lines passes through. Different values were trialled for the ingress and egress to create asymmetry.}  

We additionally trial an exponential density profile for the material:
\begin{equation}
    n(x) = \exp \left[ \frac{1-x}{h} \right],
    \label{eqn:loglin}
\end{equation}
where $h$ is the scale height of the material in units of $R_{\rm cs}$. For the ingress we consider $h=0.004$, $h=0.002$ and $h=0.001$, and for the egress $h=0.009$, $h=0.004$ and $h=0.003$. It is clear that the $\rm{N}_H(t)$ functions predicted from this density profile provide the best description of the data, as seen in Panels G and H of Figure \ref{fig:DensProf2}. For the ingress, the most consistent parameter combination is $ h = 0.002, q = 0.2$ giving $\chi^{2} / \nu = 52.0 / 44 $, and for the egress the most consistent parameter combination is $ h = 0.004, q = 0.2$ giving $\chi^{2} / \nu = 44.6/ 44 $. 

Overall, the exponential radial density profile (Equation \ref{eqn:loglin}) is most consistent with the observed $\rm{N}_H(t)$, but using the results from all density profiles, we can infer some properties of the absorbing material. The Gaussian density profile requires small fractional widths ($\Delta$), the exponential profile requires small scale heights ($h$), and the accelerating profile requires high values for $\beta$, indicating that the density of material must drop off quickly with distance from the companion star's surface in all three cases. The Gaussian and exponential profiles also highlight the asymmetry of the eclipse profiles. When using the Gaussian density profile, the parameter combination $\Delta = 0.003$ and $q = 0.4$ yields $\chi^{2} / \nu = 64.3 / 44 $ for the ingress, but yields a notably worse fit of $\chi^{2} / \nu = 277 / 44 $ for the egress. Similarly, when using the exponential density profile with $q=0.2$, a scale height of $h = 0.004$ is needed for the egress to produce $\chi^{2} / \nu = 44.6/ 44 $, while $h = 0.002$ for the ingress is required to give the comparable $\chi^{2} / \nu = 52.0 / 44 $. These results imply that more material is present in our line-of-sight during the egress than the ingress thus we consider a scenario in which the absorbing material trails behind the companion as it orbits, perhaps similar to a comet-tail. This could explain the 2.3 s asymmetry we observed in the folded eclipse profiles (Section \ref{sec:profiles}). The phase-resolved spectral fits also provide useful insights. The ingress and egress both require high ionisation parameters, with the ingress being slightly more ionised than the egress. Such levels of ionisation could be driven by irradiation from the NS itself, perhaps being collisionally ionised by a pulsar wind. The difference in ionisation and covering fraction are likely due to the orbital motion of the binary. 

%%%%%%%%%%%%%%%%%%%%%%%%%%%%%%%%%%%%%%%%%%%%%%%%%%%%%%%%%%%%%%%%%%%%%%%%%%%%%%%%%%%%%%%%%%%
% ECLIPSE MAPPING, FITS, NS MASS    
%%%%%%%%%%%%%%%%%%%%%%%%%%%%%%%%%%%%%%%%%%%%%%%%%%%%%%%%%%%%%%%%%%%%%%%%%%%%%%%%%%%%%%%%%%%
\section{Eclipse Mapping}
\label{Section:ECProfs}

\subsection{Eclipse profile model}
%\textcolor{red}{MM: Table 1 shows the results for the time-averaged spectrum but I don't see anything explicit for the out-of-eclipse spectrum}\cmtak{Perhaps this is a case of calling something by the wrong name or not explaining things fully, but I've definitely used the parameters in Table 1 as the out-of-eclipse spectrum for the eclipse profile fits.}
We represent the out-of-eclipse spectrum with the spectral model described in Section \ref{Section:SpecFit} with the parameters fixed to their best fitting values (Table \ref{tb:specpars}). We then calculate the time-dependent specific photon flux, $S(E,t)$, by multiplying the out-of-eclipse spectrum, $S_0(E)$, by an energy dependent transmission factor given by our absorption and scattering model \textsc{abssca}, which depends on ionisation parameter, covering fraction and hydrogen column density. The ionisation parameter and covering fraction are left as free model parameters, with the values for the ingress ($t<t_0$), $\xi_{\rm in}$ and $\rm{f}_{\rm cov,in}$, allowed to be different from the corresponding egress ($t>t_0$) values, $\xi_{\rm eg}$ and $\rm{f}_{\rm cov,eg}$. We calculate the column density as a function of time, $\rm{N}_H(t)$, from Equation (\ref{eqn:Nht}), which depends on mass ratio, $q$, the \textit{surface column density}, $N_{H,0}$, and an assumed parameterisation of the stellar radial density profile, $n(x)$. We trial all four forms of $n(x)$ considered in Section \ref{sec:nx}: 1) a power law with index $m$ (Equation \ref{eqn:powlaw}), 2) an accelerating wind with acceleration parameter $\beta$ (Equation \ref{eqn:beta}), 3) a Gaussian with fractional width $\Delta$ (Equation \ref{eqn:gauss}), and 4) an exponential form with scale height $h$. We allow $N_{H,0}$ and each of the $n(x)$ parameters ($m$, $\beta$, $\Delta$ or $h$, depending on the model being used) to take different values for ingress and egress. We only compute the integral in Equation (\ref{eqn:Nht}) for orbital phases with $1 \leq b(\phi) \geq x_{\rm out}$, whereas $b(\phi)<1$ corresponds to totality and $b(\phi)> x_{\rm out}$ to out-of-eclipse. In order to demonstrate the need for a layer of material around the companion star, we additionally trial a model that transitions sharply from out-of-eclipse for $b(\phi)>1$ to totality for $b(\phi)<1$. We fix the orbital period to $P=3.824$ hrs and leave the totality duration $t_e$ as a free model parameter.

The time-dependent count rate in the \text{XMM-Newton} energy range consisting of channels $I_1$ to $I_2$ is
\begin{equation}
    C(I_1,I_2,t) = \int_0^\infty~A_{\rm eff}(E,I_,I_2)~S(E,t)~dE,
\end{equation}
%\textcolor{red}{MM: the above should be between E0 and Emax?}\cmtai{No, it's formally $0$ to $\infty$. Of course in practice, $A_{\rm eff}$ is zero for $E<E_0$ and $E>E_{max}$, but the above is mathematically correct.}
where 
\begin{equation}
    A_{\rm eff}(E,I_2,I_2) = \sum_{I=I_1}^{I_2}~R_D(I,E),
\end{equation}
is the effective area of the combined instrument channels $I_1$ to $I_2$ (analogous to a photometry filter), and $R_D(I,E)$ -- the \textit{instrument response} -- is the effective area of channel $I$ (see \citealt{Rapisarda2016}). In practice, the instrument response is quantised into the response matrix stored in the rmf and arf files, which we read into our model. For convenience when fitting, both the data and model in each energy band are divided by the mean out-of-eclipse count rate such that the out-of-eclipse count rate equals $1.0$ and the totality level is $0.0$. 
%\textcolor{red}{MM: probably should say why we do this}\cmtai{It's just more convenient to not have to fit for the mean count rate in each energy band. I'm not sure of the best way to say this.}

\subsection{Results}

\begin{figure}
\includegraphics[width=1\columnwidth]{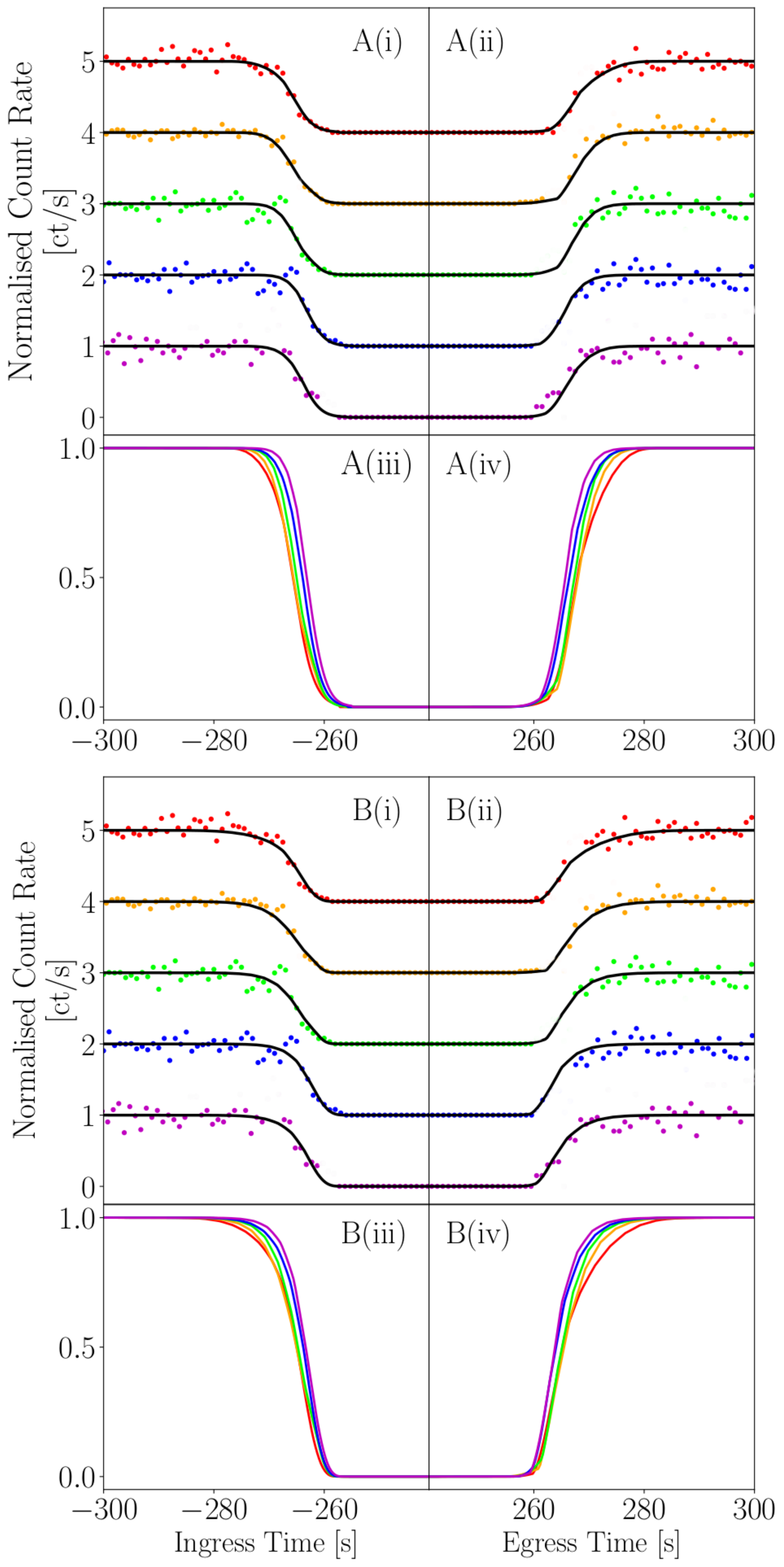} 
\vspace*{-0.5cm}
\caption{\label{fig:ECfits} Eclipse profiles resulting from fitting the eclipses of EXO 0748$-$676 in five energy bands to the eclipse profile model assuming the Gaussian density profile (panels A(i)$-$A(iv) ) and the exponential density profile (panels B(i)$-$B(iv) ). Colours are represent the 5 narrow energy bands used in our modelling and are consistent with Figure \ref{fig:ECProfs}. In each set of four panels, i and ii show the resulting fits to each individual energy band, each with a vertical offsets for visual clarity. These are $+0.0$ (magenta), $+1.0$ (blue), $+2.0$ (green), $+3.0$ (orange) and $+4.0$ (red). In each set of four panels, iii and iv show the resulting model in each energy band without a vertical offset to clearly show the energy dependent behaviour of the best-fitting eclipse profiles. Fit statistics achieved are $\chi^{2} / \nu = 822.95 / 816$ and $\chi^{2} / \nu = 816.01 / 816$ for the Gaussian (A) and exponential (B) models respectively.}
\end{figure}

\begin{figure*}
\includegraphics[width=0.95\textwidth]{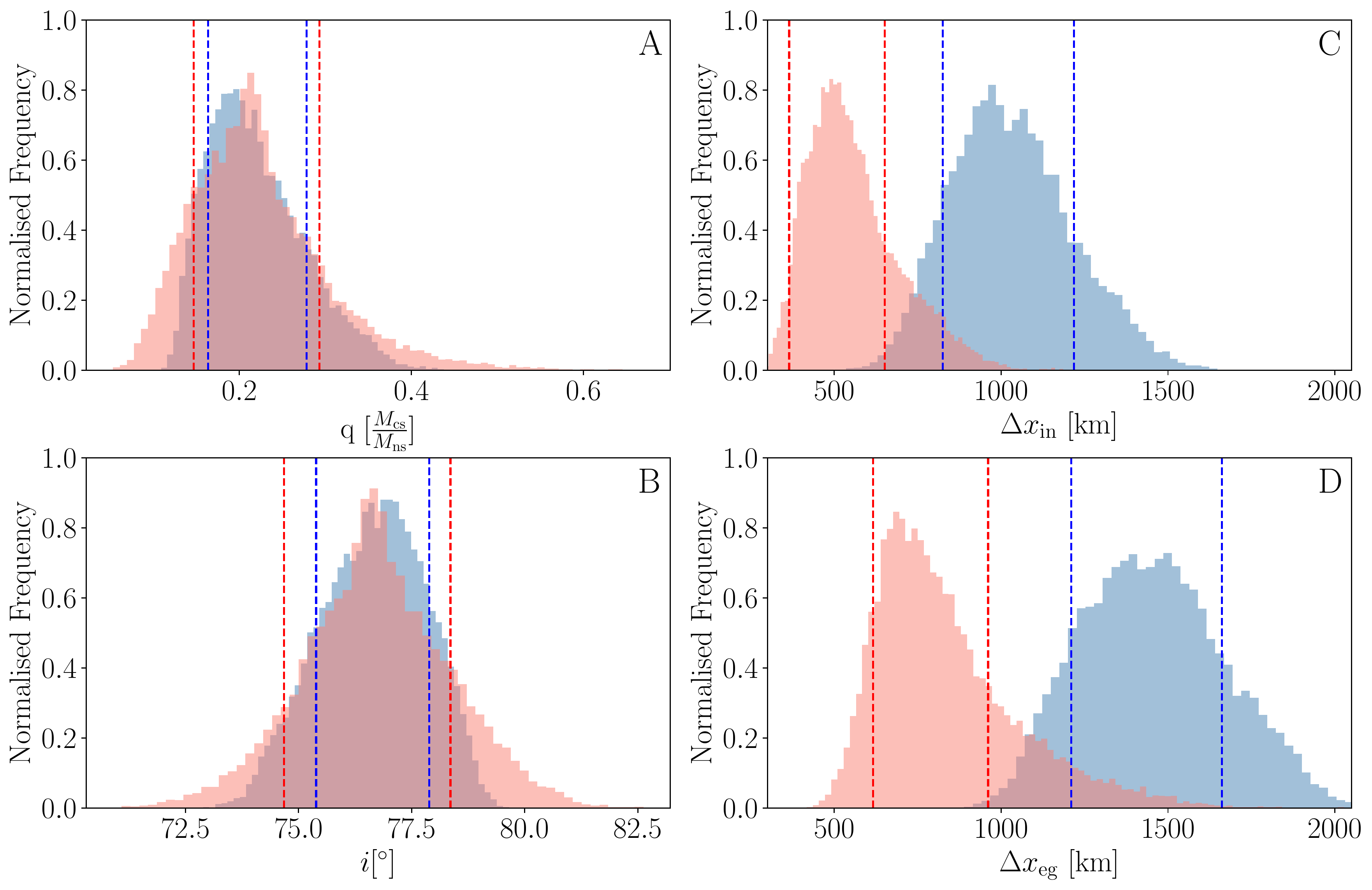}
%\vspace*{-0.5cm}
\caption{\label{fig:ECPars} Distributions of mass ratio (A), inclination angle (B), width of absorbing material during  ingress (C) and width of the absorbing material during  egress (D), from the eclipse profile model. The distributions are obtained through a Markov Chain Monte Carlo simulation with 256 walkers, 307200 steps and a burn-in of 92160 steps assuming the Gaussian density profile (blue) and the exponential density profile (red). Corresponding $1 \sigma$ confidence intervals are shown by blue and red dashed lines respectively.}
\end{figure*}

\begin{figure}
\centering
\includegraphics[width=\columnwidth]{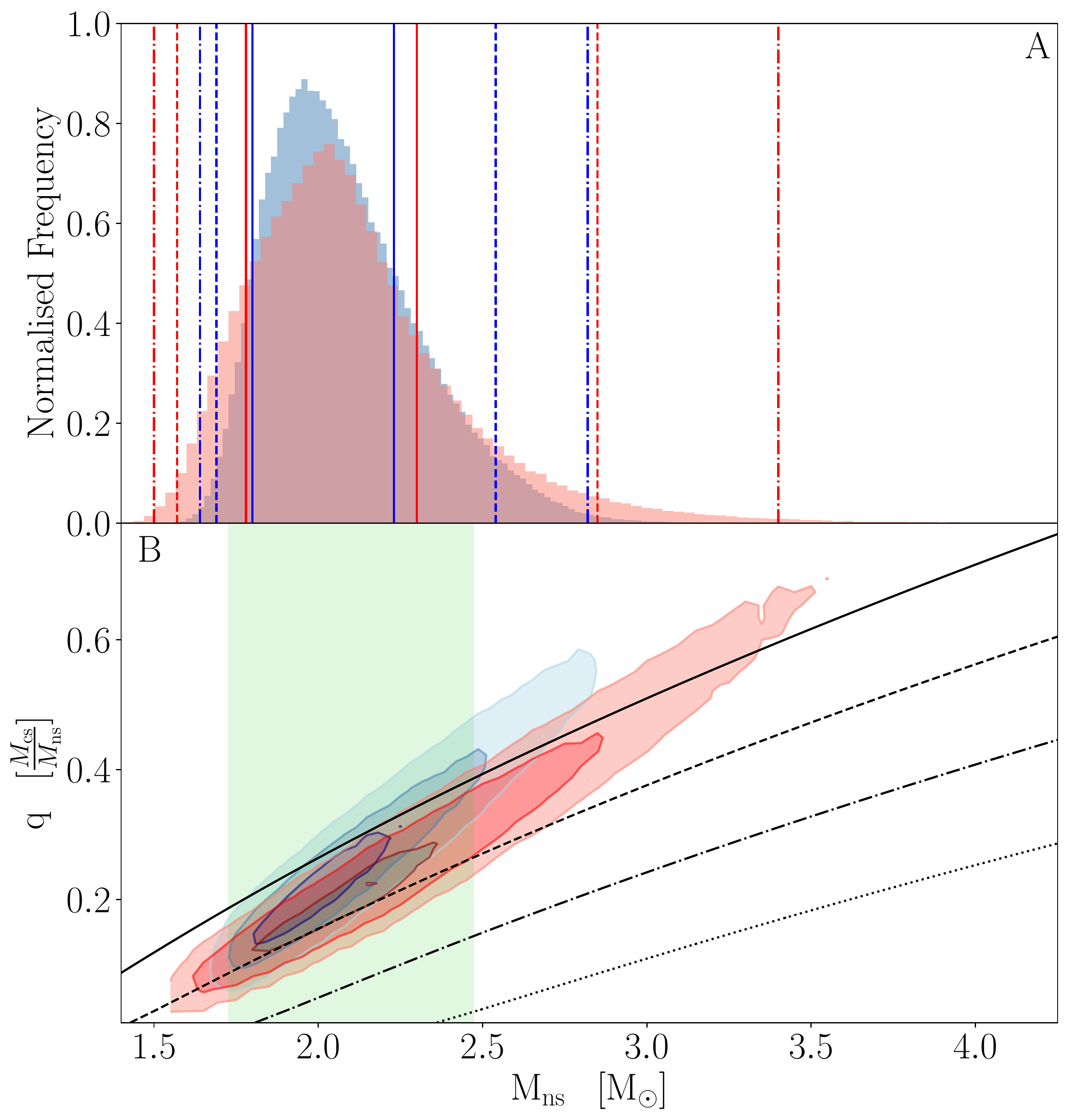} 
\caption{\label{fig:mass} Panel A: Distributions of $M_{\rm{ns}}$ assuming the Gaussian (blue) and exponential (red) density profiles within the eclipse profile model. Distributions are obtained through a Markov Chain Monte Carlo simulation with a length of 307200, 256 walkers and a burn-in length of 92160 steps. Vertical lines show $1 \sigma$ (solid), $2 \sigma$ (dashed) and $3 \sigma$ (dot-dashed) contours in blue and red, for the Gaussian and exponential density profiles respectively, which peak at $2.01 M_{\odot}$ and $2.02 M_{\odot}$. Panel B: Comparison between our measurements of $M_{\rm{ns}}$ assuming the Gaussian density profile (blue) and the exponential density profile (red), and the measurement from \citealt{Ozel2006} (green). Dark, mid and light shades of blue and red correspond to  $1$, $2$  and $3 \sigma$ contours respectively for the Gaussian and exponential density models. Black lines show the relationship between, $M_{\rm{ns}}$ and $q$, for K-corrections of 1.0 (solid), 0.9 (dashed), 0.8 (dot-dashed) and 0.7 (dotted). K-corrections closer to unity yield a lower NS mass.}
%\cmtai{I wouldn't include \citealt{MD2009} here, since they don't really \textit{measure} $q$, they just assume a range for it, and they're also using a binary mass function so their measurement is not independent of ours. It is definitely right to plot the Ozel value though -- I guess you're in the process of updating the figure to include this?} }
\end{figure}

\begin{figure*}
\includegraphics[width=1\textwidth]{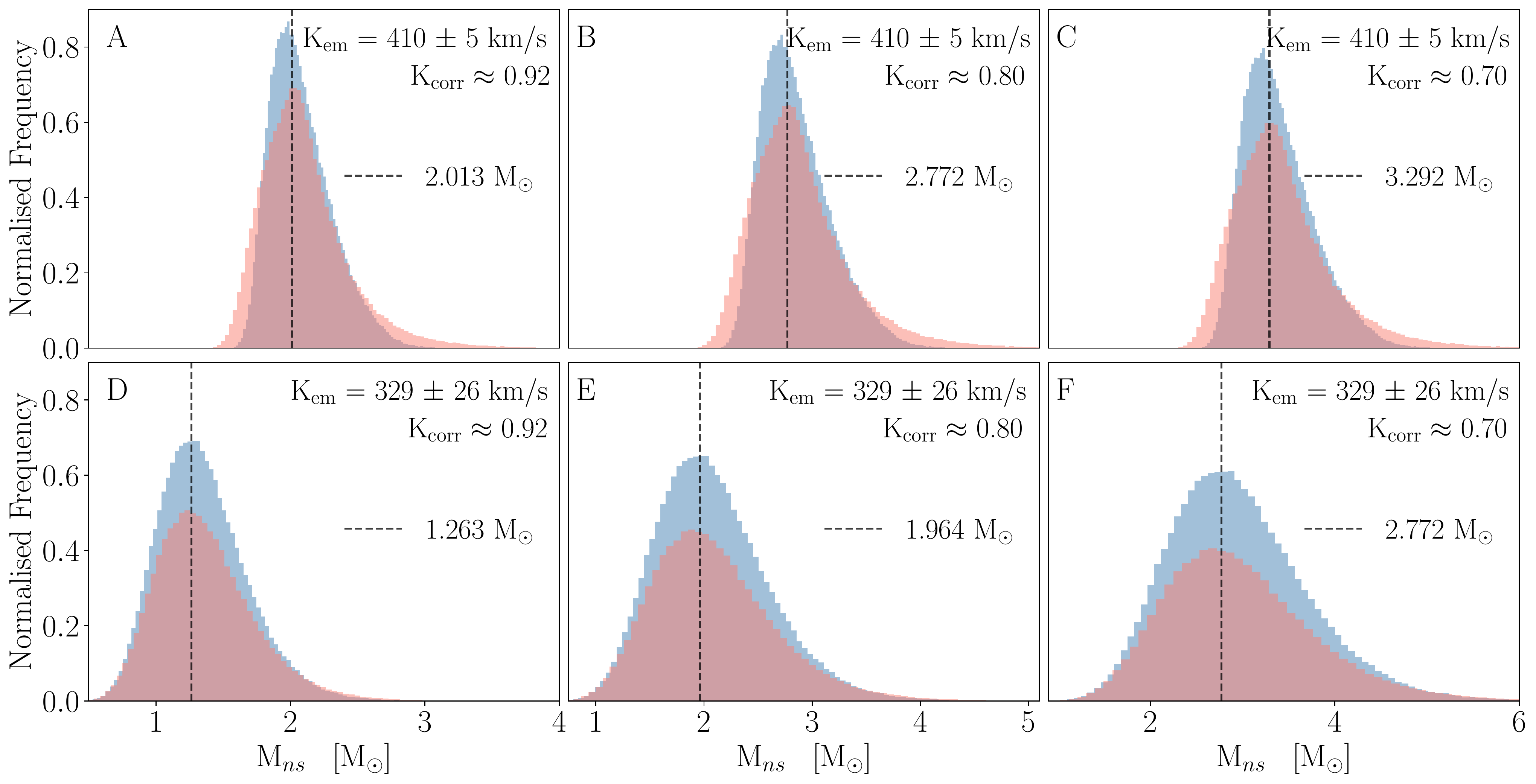}
\vspace*{-0.5cm}
\caption{\label{fig:KemComp} Distributions of $M_{\rm{ns}}$ assuming the Gaussian (blue) and exponential (red) density profiles within the eclipse profile model calculated using a different combination of $K_{\rm em}$ and K-Correction (as labelled). Also shown is a black dashed line corresponding to the peak-posterior NS mass. The K-correction in A and D corresponds to the most conservative value calculated using the mass-ratio dependent relation of \citet{MunozDarias2005} (Equation \ref{eqn:kcorr}) for our best fitting value $q = 0.222$. Column 2 (B and E) and column 3 (C and F) assume the reasonable K-corrections of 0.8 and 0.7 respectively. Emission lines considered are H$_{\alpha}$, $K_{\rm em}=$410 $\pm$ 5 km/s \citep{Bassa2009} and CIII$-$NIII blend, $K_{\rm em}=$329 $\pm$ 26 km/s \citep{Mikles2012}. We see that the peak-posterior is above the canonical NS mass unless we discard the H$_\alpha$ measurement \textit{and} employ the maximum (closest to unity) possible K-correction (panel D).}
% Distributions presented in Panel A are considered to be the most constraining while remaining conservative because the higher $K_{\rm em}$ traces a position further from the NS. Distributions presented in Panel E are considered be more realistic.

\end{figure*}

Using \texttt{xspec v12.1.1}, we simultaneously fit the eclipse profiles of EXO 0748$-$676 in the five narrow energy bands presented in Figure \ref{fig:ECProfs}B. We ignore the majority of the phase bins corresponding to out-of-eclipse and totality, leaving only a small number of bins surrounding the ingress and egress. This ensures the best-fitting parameters and statistics refer to the portions of the eclipse profiles that contain the energy-dependent behaviour. Due to the abundance of observations of EXO 0748$-$676, the eclipse duration, $t_e$ of $\sim 500$ s is well-known, but is also observed to vary \citep{Wolff2009, Parmar1991}. Therefore, we apply a Bayesian prior on $t_{\rm e}$ assuming a Gaussian peaking at $500$s with a width of $5$s.

We first trial our eclipse profile model with no absorbing material surrounding the companion. This model simply invokes a sharp transition between out of eclipse and totality. This model produces a poor fit to the observed eclipse profiles (see row 1 of Table \ref{tb:fitstats}), demonstrating the need for the layer of absorbing material. We further trial the four density profiles presented in Equations \ref{eqn:powlaw} - \ref{eqn:loglin}. Table \ref{tb:fitstats} compares the resulting fit statistics from our eclipse model for each density profile. As anticipated, the power-law density profile, with reasonable power-law indices of $m = 2$ and $m = 10$, are unable to describe the data, respectively yielding $\chi^{2} / \nu = 17700 /818$ and $\chi^{2} / \nu = 15600 /818$. Interestingly, if the power-law index is allowed to vary during the fits and be different for the ingress and egress, they rise to the high values of $m_{\rm{in}} \approx 410.0$ and $m_{\rm{eg}} \approx 760.0$. At such high values of $m$, the power-law function behaves similarly to the exponential function (Equation \ref{eqn:loglin}), so based on our previous inferences, it is not surprising that these values yield a much lower fit statistic ($\chi^{2} / \nu = 1352 /816 $). As such high indices are nonphysical, we subsequently discard the power-law density profile. The remaining three profiles yield reasonable fits to the eclipse profiles, as alluded to by the phase-resolved spectroscopy. However, we discard the accelerating wind profile because 1) the associated $\chi^{2}_{\nu} = 1.288$ is notably higher than assuming the Gaussian ($\chi^{2}_{\nu} = 1.009$) or exponential ($\chi^{2}_{\nu} = 1.000$) density profiles and 2) the associated null-hypothesis probability, $p = 10^{-39}$, indicates that this model does not reproduce the data. 

The eclipse profile model assuming the Gaussian ($\chi^{2} / \nu = 822.95 /816 $) or exponential ($\chi^{2} / \nu = 816.01 /816 $) density profiles are difficult to separate statistically, therefore, we consider both profiles going forward. The resulting eclipse profiles are shown in Figure \ref{fig:ECfits}, where panels A(i)$-$A(iv) assume the Gaussian density profile and panels B(i)$-$B(iv) assume an exponential density profile. Associated best-fitting parameters can be found in Table \ref{tb:fitpars}, with $1 \sigma$ confidence intervals obtained through a Markov-Chain Monte Carlo (MCMC) simulation (see Appendix \ref{Section:ApMCMC} for full details). The best-fitting parameters from either profile lead to a similar inference regarding the absorbing material -- the fractional widths and scale heights are both small, confirming that the material doesn't extend far out from the star's surface. Both profiles show asymmetry here, with larger fractional widths and scale heights for the egress than for the ingress. When assuming the exponential density profile, our fits suggest there is $\sim 20 \% $ more material in the egress than the ingress while assuming the Gaussian density profile increases this to $\sim 40 \%$. This further supports the hypothesis that some material is trailing behind the companion as it orbits. We find consistent ionisation parameters between the two density profiles, with both profiles finding the ingress to be more ionised than the egress, thus suggesting an incident wind or irradiation from the NS more strongly affects the ingress side of the star. The two models show variations in covering fraction, with the Gaussian model indicating the ingress side of the star is less covered by absorbing material than the egress side of the star, although the larger errors associated with $\rm{f_{cov, in}}$ mean that the two sides could have similar coverings. In comparison, the exponential model shows the covering fractions to be consistent between the two sides of the star. The best-fitting mass ratio, $q$, and inclination angle, $i$, are found to be independent of the assumed radial density profile; we, therefore, have increased confidence in these values and our subsequent inference of the NS mass. Nonetheless, the energy-dependent eclipse timings strongly depend on the chosen density profile, so it is necessary to understand the density of the material to model the features of the eclipses caused by absorption. When comparing the eclipses modelled with the Gaussian density profile to the eclipses modelled with the exponential density profile (i.e. comparing panels A(iii) and A(iv) with panels B(iii) and B(iv) of Figure \ref{fig:ECfits}), we see the energy bands are more dispersed at the start ($\rm{t_{10,in}}$) and end ($\rm{t_{10,eg}}$) of totality when modelled with the Gaussian density profile. As a result, the Gaussian model does not capture the energy independence of the observed $\rm{t_{10}}$ times as well as the exponential model. Both models are capable of reproducing the observed $\rm{t_{90,in}}$ and $\rm{t_{90,eg}}$. 

\begin{table}
\begin{center}
\begin{tabular}{ |c|c|c|c|c|c|c| } 
\hline
Density Profile & Parameter(s) & $\chi^{2}$ & $\nu$ & $\chi^{2}_{\nu}$ & $p$ \\
\hline
\hline
 No Material & - & 4546.74 & 818 & 5.56 & $10^{-256}$\\ 
 \hline
Power-law & m\ =\ 2.00 * & 17700 & 818 & 21.61 \\ 
 & m\ =\ 10.0 * & 15400 & 818 & 18.81\\ 
 & $\rm{m_{in}}$ \ =\ 413.6 & 1352.1 & 816 & 1.65 & $10^{-167}$\\ 
  & $\rm{m_{eg}}$ \ =\ 764.9 &  & & \\
\hline
Accelerating & $\beta_{\rm{in}}$\ =\ 5.48 & 1050.6 & 816 & 1.288 & $10^{-39}$\\ 
 & $\beta_{\rm{eg}}$\ =\ 6.20 & & \\ 
\hline
Gaussian & $\Delta_{\rm{in}}$\ =\ 0.0035 & 822.95 & 816 & 1.009 & 0.581 \\ 
 & $\Delta_{\rm{eg}}$\ =\ 0.0047 & & \\ 
\hline
Exponential & $\rm{h}_{\rm{in}}$\ =\ 0.0023 & 816.01 & 816 & 1.000 & 0.610 \\ 
 & $\rm{h}_{\rm{eg}}$\ =\ 0.0027 & & \\ 
\hline \\
\end{tabular}
\end{center}
\caption{\label{tb:fitstats} Fit statistics and characteristic density profile parameters obtained from fitting the eclipse mapping model simultaneously to eclipses profiles of EXO 0748$-$676 in five energy bands, for each assumed density profile. From left to right: the assumed density profile of the absorbing material, key parameters governing the density profile, chi-squared, number of degrees-of-freedom, reduced chi-squared 
%\textcolor{red}{MM: you don't strictly need this as it's only one column divided by the other} and null-hypothesis probability. 
The key density parameters are power-law index, $m$, acceleration parameter, $\beta$, fractional width of the material, $\Delta$ and scale height, $h$, for the power-law, accelerating, Gaussian and exponential density profiles respectively. * Parameter fixed for the duration of the fit.}
\end{table}

\begin{table}
\begin{center}
\begin{tabular}{|c|c|c|}
\hline
Parameter & Gaussian Density Profile & Derived Values\\
\hline
\hline
$t_{\rm e}$ & 503.21 $\pm ^{1.62}_{1.48}$ & $i = 76.52 \pm ^{1.37}_{1.13}$\\
\hline
$q$ & 0.221 $\pm ^{0.057}_{0.057}$ & $M_{\rm{ns}} = 2.01  \pm^{0.22}_{0.21}$ \\
\hline
$N_{H0, \rm{in}}$ & 2364.32 $\pm _{666.8}^{682.7}$ &\\
\hline
$N_{H0, \rm{eg}}$ & 4842.15 $\pm _{543.3}^{506.2}$ &\\
\hline
$\log(\xi)_{\rm{in}}$ & 3.49 $\pm _{0.084}^{0.460}$ &\\
\hline
$\log(\xi)_{\rm{eg}}$ & 2.82 $\pm _{0.043}^{0.009}$ &  \\
\hline
$\rm{f_{cov, in}}$ & 0.897 $\pm _{0.062}^{0.064}$ & \\
\hline
$\rm{f_{cov, eg}}$ & 0.995 $\pm _{0.002}^{0.002}$ &\\
\hline
$\Delta_{\rm{in}}$ & 0.0035 $\pm _{0.0006}^{0.0006}$ & $\Delta x = 1033 \pm ^{194}_{186}$ km \\
\hline
$\Delta_{\rm{eg}}$ & 0.0049 $\pm _{0.0007}^{0.0008}$ & $\Delta x = 1446 \pm ^{216}_{236}$ km \\
\hline
\\
\\
\hline
Parameter & Exponential Density Profile & Derived Values \\
\hline
\hline
$t_e$ & 504.11 $\pm _{0.62}^{0.56}$ & $i = 76.47 \pm ^{1.89}_{1.79}$\\
\hline
$q$ & 0.222 $\pm _{0.075}^{0.070}$ & $M_{\rm{ns}} = 2.02 \pm^{0.29}_{0.27}$\\
\hline
$N_{H0, \rm{in}}$ & 3819.7 $\pm _{1040.2}^{1023.9}$ &\\
\hline
$N_{H0, \rm{eg}}$ & 4320.3 $\pm _{972.4}^{881.3}$ &\\
\hline
$\log(\xi)_{\rm{in}}$ & 3.30 $\pm_{0.013}^{0.349}$ &\\
\hline
$\log(\xi)_{\rm{eg}}$ & 2.76 $\pm _{0.125}^{0.081}$ &\\
\hline
$\rm{f_{cov, in}}$ & 0.998 $\pm_{0.001}^{0.002}$ & \\
\hline
$\rm{f_{cov, eg}}$ & 0.996 $\pm_{0.009}^{0.005}$ &\\
\hline
$\rm{h_{in}}$ & 0.0023 $\pm_{0.0001}^{0.0012}$ & $\Delta x = 679 \pm ^{222}_{107}$ km\\
\hline
$\rm{h_{eg}}$& 0.0027 $\pm_{0.0009}^{0.0004}$ & $\Delta x = 797 \pm ^{104}_{251}$ km \\
\end{tabular}
\end{center}
\caption{\label{tb:fitpars} Best fitting parameters obtained when fitting the eclipse mapping model simultaneously to the eclipse profiles of EXO 0748$-$676 in five energy bands. We implement both the Gaussian (top table) and exponential (bottom table) density profiles for the absorbing material around the companion star. From top to bottom the model parameters are totality duration, mass ratio, surface column density for the ingress and egress, ionisation parameter for the ingress and egress, covering fraction for the ingress and egress. The final two parameters are the fractional widths of the surrounding material layer for the ingress and egress in the case of the Gaussian density profile and the surrounding material's scale height for the ingress and egress in the case of the exponential density profile. The third column details values that have been derived from the model parameters in that row. Here, $M_{\rm{ns}}$ is the NS's mass, $i$ is the binary inclination angle in degrees and $\Delta {\rm x}$ is the physical width of absorbing material from the companion star's surface.}
\end{table}

% 4.3 
\subsection{Neutron Star Mass}
We can calculate the NS mass from the binary mass function by rearranging Equation (\ref{eqn:f}) to get
\begin{equation}
    M_{\rm ns} = \frac{P K_{\rm em}^3}{2\pi G} \frac{1}{(K_{\rm em}/K)^3} \frac{(1+q)^2}{\sin^3 i},
    \label{eqn:massfunc}
\end{equation}
where $K_{\rm em}$ is the semi-major amplitude of the radial velocity curve of the observed stellar emission lines. Since emission lines originate from the irradiated face of the companion, they do not trace the centre of mass of the companion, but instead a region of the companion that is closer to the centre of mass of the binary. Therefore $K_{\rm em} < K$, and a \textit{K-correction} $(K_{\rm em}/K) < 1$, is required in order to infer $K$ from $K_{\rm em}$. \citet{MunozDarias2005} derive physical upper and lower limits for the K-correction as a function of mass ratio. The upper limit ($K_{\rm em}$ closest to $K$) corresponds to the emission line originating entirely from the point on the companion's Roche-Lobe surface that is furthest from the NS whilst still being visible to it, giving
\begin{equation}
    K_{\rm em}/K = 1 - 0.2134 q^{2/3} (1+q)^{1/3}.
    \label{eqn:kcorr}
\end{equation}
The lower limit corresponds to the line originating entirely from the L1 point. In reality, most systems fall in between these two limits. 

Different emission lines that originate predominantly from different parts of the irradiated face of the companion star will produce different measured $K_{\rm{em}}$ and will disappear at orbital phases $\sim 0$ (i.e. during or close to eclipse). Higher $K_{\rm{em}}$ values correspond to lines emitted furthest from the NS and therefore will have a K-correction closer to unity. For EXO 0748$-$676 there have been four independent studies resulting in radial velocity measurements from five emission features, all of which appear to be associated with the companion star. In outburst \citet{MD2009} obtain 310 $\pm$ 10 km/s from He${\rm{II}}$ and \citet{Mikles2012} find 329 $\pm$ 26 km/s from the Bowen blend (C${\rm{III}}$ - N${\rm{III}}$).  In quiescence \citet{Bassa2009} used a Doppler mapping (DM) technique to obtain 345 $\pm$ 5 km/s and 410 $\pm$ 5 km/s respectively for He${\rm{II}}$ and H$_{\alpha}$. By applying a Gaussian fitting technique to the same H$_{\alpha}$ emission, \citet{Bassa2009} find $333 \pm 5$ km/s. The authors favour the DM derived values since this technique does not require symmetric line profiles, and consider the Gaussian model to be too simplistic to reproduce the complex shape of the H$_{\alpha}$ emission. Finally, \citet{Ratti2012} found 308.5 $\pm$ 3.9 km/s from a weighted average of H$_{\beta}$ and H$_{\gamma}$. 

We first chose to implement the H$_{\alpha}$ velocity measurement of $K_{\rm{em}} = 410 \pm 5$ km/s from \citet{Bassa2009} because it is the largest value, corresponding to a line emitted further from the NS than any other detected lines and therefore has a K-correction closer to unity ($K_{\rm{em}} $ closest to K). For a conservative approach, we use the upper limit of the K-correction (Equation \ref{eqn:kcorr}), meaning that our mass measurement will be an under-estimate. We use the mass ratio measured from our eclipse profile fits, implementing the results of both the Gaussian and exponential models. For each model, we obtain a posterior probability distribution for $M_{\rm ns}$ by running an MCMC simulation with 256 walkers for a total of 307200 steps after a burn-in period of 92160 steps (see Appendix \ref{Section:ApMCMC} for full details). For each step in the chain, we calculate $i$ from $q$ and $t_e$ (Equation \ref{eqn:inc}), draw a value of $K_{\rm{em}}$ from a Gaussian distribution with centroid $410$ km/s and width $5$ km/s, and finally calculate $M_{\rm ns}$ from Equations (\ref{eqn:massfunc}) and (\ref{eqn:kcorr}). Figure \ref{fig:ECPars} shows the resulting posterior distributions of mass ratio, inclination and surrounding material width for the ingress and egress. Figure \ref{fig:mass}A shows the resulting $M_{\rm ns}$ distribution, with $1$ (solid), $2$ (dashed) and $3~\sigma$ (dot-dashed) confidence levels. For both figures \ref{fig:ECPars} and \ref{fig:mass}, the Gaussian and exponential models are coloured blue and red respectively.

We see that the inferred NS mass is independent of the assumed density profile. We find $M_{\rm ns} = 2.01 \pm ^{0.22}_{0.21} M_{\odot}$ and $M_{\rm ns} = 2.02 \pm ^{0.29}_{0.27} M_{\odot}$ when we model the absorbing material with a Gaussian or exponential density function respectively. Note that for both models, the canonical NS mass of $1.4~M_{\odot}$ falls outside of the $3 \sigma$ contours. We infer this high value for the mass despite employing the most conservative possible K-correction (Equation \ref{eqn:kcorr}). Applying a more realistic K-correction would increase the peak posterior mass values. Figure \ref{fig:mass}B demonstrates how our mass measurement depends on the mass ratio and K-correction. Solid, dashed, dot-dashed and dotted lines correspond to $K_{\rm em}/K =~1,~0.9,~0.8$ and $0.7$ (for comparison, Equation \ref{eqn:kcorr} gives $K_{\rm {em}} / K \approx 0.92$ for $q=0.2$). We see that employing a reasonable value of $K_{\rm em}/K = 0.8$ (see Figure 4 of \citealt{MunozDarias2005}) pushes the mass measurement to $M_{\rm ns} \approx 2.8~M_\odot$ (see Figure \ref{fig:KemComp}A-C), which is more massive than the most massive confirmed NS to date ($\sim 2.1~M_\odot$: \citealt{Cromartie2019}). The distributions can also extend into the $3.0 - 5.0 M_{\odot}$ compact object mass-gap, where it becomes uncertain if the compact object would be a NS or a BH. Since the primary in EXO 0748$-$676 is confirmed to be a NS by the presence of Type 1 busts with associated burst oscillations \citep{Ozel2006, Galloway2010}, future direct measurements of $K$ may even inform on the observational lower bound of the mass gap. Furthermore, our consistency with the mass measurement presented by \citet{Ozel2006} ($M_{\rm ns} = 2.10 \pm 0.28 M_{\odot}$ and $r_{\rm ns} = 13.8 \pm 1.8$km) is encouraging for boths their PRE burst method and our eclipse profile model. Our findings improve confidence in their conclusion, agreeing that a harder EoS is required for nuclear matter.

The difference in each measured $K_{\rm{em}}$ arises because each line will trace a different spatial zone between the irradiated face of the companion star and the L1 point, and since the highest value H$_{\alpha}$ = 410 $\pm$ 5 km/s \citep{Bassa2009} is thought to originate closer to the companion, it is reasonable to assume that this line is the most constraining. However, it is important to consider the caveats associated with this particular K$_{\rm{em}}$ measurement. Firstly, the approach taken by \citet{Bassa2009} is more typical of outburst studies, when the lines are assumed to be formed by irradiation. At the time EXO 0748$-$676 had just entered quiescence, making the detection of H$_{\alpha}$ unusual. Furthermore, the author's favoured result of $K_{\rm{em}}$ = 410 $\pm$ 5 km/s obtained via a well-tested but indirect DM technique differs substantially from their result of $K_{\rm{em}}$ = 333 $\pm$ 5 km/s, obtained directly from Gaussian fits to the same emission line. While \citet{Bassa2009} justify their preference for the velocities obtained via DM by suggesting that the DM method accounts for the shape of the emission line not being Gaussian, we highlight that $K_{\rm{em}}$ = 410 $\pm$ 5 km/s is inconsistent with all other radial velocity measurements for this source, including the other values presented by \citet{Bassa2009}. Additionally, our conservative approach towards the K-correction (i.e. $K_{\rm{em}}$ / $K \sim$ 1) implies that the emitting area of the companion star is very small which is difficult to reconcile with the strong H$_{\alpha}$ emission component that is observed \citep{Bassa2009}. As shown in Figures \ref{fig:mass} and \ref{fig:KemComp}, a more reasonable K-correction easily increases the peak of the distributions to $> 3M_{\odot}$. These high masses are inconsistent with a number of possible EoS and are substantially higher than any observed NS mass to date. 

Another problem with assuming $K_{\rm{em}}$ = 410 $\pm$ 5 km/s is the small error ($\sim 1$ per cent). Underestimated uncertainties are often a problem associated with DM techniques, and in the case of \citet{Bassa2009}, the error is assigned from the variation in the centroid velocity of the large spot seen in the DM, but only considers the effect of using the wrong systematic velocity. Recently, \citet{Wang2017} and \citet{Jimenez-Ibarra2018} tackled the issues with the errors in DM spots, using a newly developed code that computes the error using bootstraps DMs, obtaining more realistic errors on the radial velocities of $6 - 8$ per cent from very significant DM spots. We note that if $K_{\rm{em}}$ = 410 $\pm$ 5 km/s \citet{Bassa2009} is trusted, the error is likely more significant than the one quoted and therefore, may not rule out the canonical NS mass. We also note that assuming $K_{\rm{em}}$= 410 $\pm$ 5 km/s violates the assumption that the disk rim orbits at Keplerian velocities \citep{Mikles2012}. This does not provide a definitive reason to discard H$_{\alpha}$ because sub-Keplerian disk rim velocities have previously been measured \citep{Somero2012}, but when combined with the other caveats presented here, provides reason to consider the other measured $K_{\rm{em}}$ values.

We additionally consider $K_{\rm{em}}$= 329 $\pm$ 26 km/s from \citet{Mikles2012}. This measurement is robust, using a standard technique, and the value is consistent with all others except for the DM H$_{\alpha}$ measurement from \citet{Bassa2009}. Note that this value is consistent with their H$_{\alpha}$ velocity derived via Gaussian fitting. Figure \ref{fig:KemComp} shows how our mass posterior changes when we consider these two different $K_{\rm em}$ values for three different values of the K-correction (as labelled). Here K$_{\rm{corr}} \approx 0.92$ corresponds to the conservative, mass ratio-dependent K-correction calculated using our MCMC simulations and Equation \ref{eqn:kcorr}. The K-corrections used in Panels B, C, E and F are found by drawing a value from Gaussian distributions peaking at $0.80$ and $0.70$ respectively with widths of $0.025$. We chose this Gaussian width as it's comparable to the width of the calculated K-correction distribution peaking at $\sim 0.92$. For the larger $K_{\rm{em}}$ (panels A-C), the distribution peaks at $M_{\rm ns} \sim 2~M_{\odot}$ even for the largest possible K-correction value (panel A), and the canonical NS mass is ruled out with $>3\sigma$ confidence. For the smaller $K_{\rm em}$ (panels D-F), the distribution peaks at $M_{\rm ns} \sim 2~M_{\odot}$ for the most realistic K-correction ($K_{\rm corr} = 0.8$: panel E). For the Gaussian and exponential model respectively these distributions peak at $M_{\rm ns} = 1.95 \pm ^{0.60}_{0.50} M_{\odot}$ and $M_{\rm ns} = 1.97 \pm ^{0.53}_{0.49} M_{\odot}$. However, the distribution is broader due to the larger error bar on the $K_{\rm em}$ measurement, meaning that the canonical NS mass is not ruled out. Moreover, pushing the K-correction to its highest possible value moves the peak to $M_{\rm ns} \sim 1.3~M_\odot$ (panel D). As such, our modelling favours a $M_{\rm ns} \gtrsim 2~M_\odot$ NS, but does rule out the canonical value of $M_{\rm ns} \sim 1.4~M_\odot$ when considering the robust radial velocity amplitude measurement from the Bowen blend.

\section{Discussion}
\label{Section:Discussion}
We have modelled the energy-dependent X-ray eclipse profiles of EXO 0748$-$676 from which we have inferred an inclination of $i \sim 77^\circ$, mass ratio $q \sim 0.2$ and thus NS mass $M_{\rm ns} \gtrsim 2~M_\odot$. We infer the presence of a narrow ($\sim 500-1500$ km) region of ionised material around the low mass companion star which absorbs soft X-rays more efficiently than hard X-rays. The presence of such a region explains the energy dependence of the extended ingress and egress profiles. In particular, ingress and egress are longer for softer X-rays, but the start and end times of totality are more or less independent of energy. The egress is $\sim 2.3$ s longer than the ingress. This can be explained if the absorbing material trails slightly behind the companion star during its orbit. Our fits require the layer of material to be $\sim 20-40$ per cent thicker behind the star than it is in the direction of orbital motion. The spectroscopic mass of the M-dwarf companion star is estimated to be $\sim 0.45 M_{\odot}$ \citep{Parmar1986}, assuming it to be on the main sequence; and our eclipse mapping analysis returns a consistent value of $M_{\rm cs} \sim 0.44~M_\odot$. Such a star would not typically launch its own wind, therefore we explore other possible origins for the absorbing material.

Rapidly rotating M-dwarf stars are known to exhibit strong magnetic fields, of the order of a few kilo-Gaus (kG) \citep{Johns-Krull1996, Shulyak2019, Kochukhov2021}. For a tidally locked binary system, the secondary star will have a rotation period equal to the binary period, which for EXO 0748$-$676 is a few hours. Therefore, it is possible that the companion star could have a sufficiently strong magnetic field to induce a \textit{slingshot prominence} \citep{CollierCameron1991,CollierCameron1996,Steeghs1996,Ferreira2000}. This occurs when strong, active magnetic regions on the surface of the star interact with the forces of rotation which pull and distort the magnetic field lines from the star \citep{Steeghs1996}. The magnetic field lines loop out from the surface of the star, typically near the equator and carry stellar material along them. This could introduce absorbing material into our line-of-sight, but only for an $\sim$edge-on system. For our preferred inclination of $i \sim 77^\circ$, our sight-line would miss such an equatorial prominence, and so we disfavour this interpretation.

The interaction between incident radiation from the NS, or a pulsar wind, and any outflow (not accretion flow) from the companion star are known to cause intra-binary shocks between the binary components \citep{An2018}. If an intra-binary shock is present, it could produce the necessary absorption. Here, the bow shock would channel any ionised stellar material into a parabolic shape around the companion star, causing extra absorption to occur close to the companion star. This scenario is well-motivated by observations with its effects being seen in pulse profiles \citep{An2018, Polzin2020, MiravalZanon2021}. We do not entirely rule out this possibility but cannot comment further as the complexities associated with modelling an intra-binary shock are beyond the scope of this paper. Note, however, that this scenario may not explain why the absorbing region is so narrow, as required by the data.

Our eclipse profile modelling measures $M_{\rm{cs}} \sim 0.44 M_{\odot}$ and $R_{\rm{cs}} \sim 300,000$~km ($0.43R_\odot$), which are consistent with known mass and radius values of M2V - M3V main sequence stars \citep{Kaltenegger2008}, thus supporting \citet{Parmar1986}, who suggested the companion is a main sequence M-dwarf. However, no absorption lines have ever been observed from the companion star in EXO 0748$-$676 and so we do not dismiss the possibility that the companion may not be a main sequence star, as is suggested by \citet{Mikles2012}. In this case, we may simply be seeing the expanded outer layers of the star as it evolves off the main sequence, or excess material from Roche lobe overflow \citep{Pols1998}. This alternative scenario is plausible, particularly if the star is within the short-lived sub-giant phase. However, subsequent giant-branch phases can see the radius of low-mass stars increase by $100 - 10,000$ times ($L_{\rm{cs}} \propto R_{\rm{cs}}^{2}$) \citep{Pols1}, thus requiring the absorbing material to extend further from the stellar surface than the narrow layer of material we infer. 
%\textcolor{red}{MM: have we checked if you can even get a MS star in this tight an orbital configuration without invoking a random capture?} \textcolor{red}{MM: we should add a ref for this or expand on this point}

The X-ray eclipse profiles observed here are reminiscent of the radio eclipse profiles observed for spider pulsars, which also feature extended, frequency-dependent eclipse profiles and egress/ingress asymmetry \citep{Fruchter1988, Polzin2018}. In these systems, incident radiation from the NS bombards and heats the outer layers of the companion star, resulting in their ablation from the surface of the companion. This process liberates and ionises material from the stellar surface, encasing the companion in a region of highly ionised material that trails somewhat behind the star due to the binary's orbital motion \citep{Fruchter1990, Polzin2018}. Therefore, given the similarities between the observations of eclipsing spider pulsars and those we present here, we consider the possibility that EXO 0748$-$676 is a progenitor to these pulsar systems. We find that the absorbing material around the companion star has a steep radial density profile and a characteristic width of only a few per cent of the companion star's radius, therefore, we invoke an \textit{early ablation} scenario in which incident radiation from the NS or a pulsar wind has just started to ablate the outer layers of the companion. For EXO 0748$-$676 it is possible that kinetic energy from the disk wind \citep{Ponti2014} is itself, sufficient to cause ablation or a contributing source of incident radiation that subsequently leads to ablation. Since EXO 0748$-$676 existed in an accretion powered state for more than two decades, it is reasonable to consider that the disk wind may have contributed to the presence of the absorbing material. Assuming ablation to be driven by thermalisation, the response is the thermal (Kelvin-Helmholtz) timescale of the atmosphere ($\sim 10^{12}$ yrs). This is consistent with the lifetime of a $0.4 M_{\odot}$ M-star \citep{Pols1}. Therefore, in the case of our early ablation scenario either not much material will have been ablated and/or the material has not had long enough to diffuse away from the star's surface. Both could explain why the absorbing region is so narrow. The highly ionised material inferred from our analysis ($2.8 \lesssim \log\xi \lesssim 3.5$) is consistent with it originating from an ablation process caused by irradiation from the NS (and accretion flow), and/or a pulsar wind \citep{Fruchter1990, MiravalZanon2021}, as is thought to be the case for spider pulsars. 

Similar interpretations of EXO 0748$-$676 being a spider progenitor have been suggested by \citet{Ratti2012}, who performed phase-resolved optical spectroscopy of the companion star's emission lines, and \citet{Parikh2021}, who analysed UV spectroscopic data and quasi-simultaneous Swift X-ray observations of the source in quiescence. \citet{Ratti2012} attributed the broad emission lines to an outflow driven by a pulsar wind and/or X-ray heating from the stellar surface, since broadening via tidally-locked rotation alone would indicate an unfeasibly heavy NS ($M_{\rm ns} \gtrsim 3.5~M_\odot$, which though unlikely, is not entirely ruled out by our analysis). These authors noted that this scenario could also explain the lack of observed disk emission features and the observed variability in the g-band light curve. \citet{Parikh2021} notes that the broadening of the C IV line observed in EXO 0748$-$676 is similar to that of UV lines observed in PSR J1023+0038 (a known tMSP), which are thought to be broadened by a stellar outflow driven by the pulsar wind. Our results provide further evidence that EXO 0748$-$676 could be a \textit{transitional redback pulsar}; i.e. a progenitor to a spider pulsar. Under this hypothesis, EXO 0748$-$676 will eventually transition to a redback pulsar, during which ablation will continue until it finally evolves into an isolated millisecond radio pulsar. However, to date, no radio pulsations have been observed from this object so evidence for our interpretation remains circumstantial.

Similarities between the observed behaviour of EXO 0748$-$676 and the evaporation or disintegration of exoplanet atmospheres further suggest that the companion star in EXO 0748$-$676 is undergoing ablation. In the context of planetary ablation, it is usual to consider the photo-thermal escape flow cycle \citep{Moore2007}. When incident radiation deposits energy into the geosphere, the outer layers of the atmosphere expand. This is known as \textit{upflow}. With sufficient energy, the upflow becomes a material outflow, that can escape gravity but remains trapped by the planet's magnetosphere. In this context, ablation is then the process describing out-flowing material that has escaped the magnetospheric boundary \citep{Moore2007}. Any liberated atmospheric material then trails behind the exoplanet as it orbits, causing ingress/egress asymmetry in the observed transits \citep{LecavelierDesEtangs2010, Vanderburg2015}. Typically, this is seen in systems where the planet is in a tight orbit with its host star, so a similar ablation process would, in theory, apply to short period XRBs. 

Interestingly, a similar evaporative wind scenario is considered by \citet{Parmar1991} to explain the heavily extended ingress and egress durations in EXO 0748$-$676 as observed by \textit{EXOSAT} between $1985$ and $1989$. The ingress and egress durations are found to be as long as $40$ s, are highly variable and show significant asymmetry. Since the atmospheric scale height for a star of the same type as the companion is $\sim 100$ km \citep{Parmar1986, Parmar1991}, they conclude that the scale height must be enhanced, possibly by an X-ray induced evaporative wind, to explain the long durations and their variability. Our analysis infers a material layer $4 - 20$ times larger than their calculated scale height, supporting the notation that an extended atmosphere or layer of ablated material must be surrounding the companion and significantly extending the ingress and egress duration. 

The early ablation scenario and the hypothesis that EXO 0748$-$676 is a redback tMSP can be tested through extended observations of this source. While currently in quiescence, the source should be rotationally powered, allowing us to search for radio and gamma-ray pulsations. The timescales of transitional pulsar phases are not well established, so only extended, regular monitoring of EXO 0748$-$676 can confirm if it will transition back to an accretion powered state. Similar monitoring will allow us to study the suspected ablation process and the predicted evolution of this source towards a black widow pulsar. 

We combine the mass ratio and binary inclination inferred from our eclipse mapping analysis with the previously measured binary mass function \citep{MD2009, Bassa2009, Mikles2012, Ratti2012} to yield an estimate for the NS mass. When implementing the H$_{\alpha}$ radial velocity amplitude of $K_{\rm em}=$410 $\pm$ 5 km/s derived via DM \citep{Bassa2009}, we find $M_{\rm ns} = 2.01 \pm ^{0.22}_{0.21} M_{\odot}$ and $M_{\rm ns} = 2.02 \pm ^{0.29}_{0.27} M_{\odot}$ when we model the absorbing material with a Gaussian or exponential density function respectively. Under these conditions, we find that the canonical value of $M_{\rm ns} = 1.4 M_{\odot}$ falls outside of the $3 \sigma$ contour for both models. For these measurements, we used a {\it lower limit} of the binary mass function, which was inferred from emission lines that originate from the irradiated face of the companion star \citep{Bassa2009, Mikles2012}. The true binary mass function is calculated by applying a K-correction \citep{MunozDarias2005}. We have considered the most conservative (i.e. closest to unity) K-correction possible when implementing the radial velocity of H$_{\alpha}$, such that a more realistic K-correction would \textit{increase} the inferred mass. Indeed, a realistic K-correction of $\sim 0.7-0.8$ yields mass distributions that extend into the $3.0 - 5.0 M_{\odot}$ compact object mass-gap. 

However, the detection H$_{\alpha}$ during quiescence is atypical and there are a number of caveats that require consideration before confidently implementing the DM derived H$_{\alpha}$ velocity of 410 $\pm$ 5 km/s \citep{Bassa2009}. Crucial considerations are its inconsistency with all other $K_{\rm em}$ measurements for this source, including the H$_{\alpha}$ velocity measurement of $K_{\rm em}$ = 345 $\pm$ 5 km/s derived from direct Gaussian fits to the emission line rather than DM \citep{Bassa2009}, and the small associated error ($\sim$ 1 per cent). We therefore also implement the Bowen blend radial velocity amplitude of $K_{\rm em}=$329 $\pm$ 26 km/s \citep{Mikles2012} to explore the limits we get without relying on H$_\alpha$. In this case (i.e. discarding the H$_\alpha$ measurement), assuming a realistic K-correction of $K_{\rm corr}=0.8$ \citep{MunozDarias2005} gives a NS mass of $M_{\rm ns} = 1.95 \pm ^{0.60}_{0.50} M_{\odot}$ and $M_{\rm ns} = 1.97 \pm ^{0.53}_{0.49} M_{\odot}$ for Gaussian and exponential radial density functions respectively. However, these distributions are broader and only rule out the canonical mass at $1 \sigma$ level, and assuming the maximum possible K-correction yields a peak-posterior below the canonical mass, although, the maximum possible K-correction is somewhat nonphysical. We therefore favour a NS mass $\gtrsim 2~M_\odot$, but cannot definitively rule out the canonical $\sim 1.4~M_\odot$ value. Future radial velocity measurements of lines originating directly from the companion star surface would dispense with the need for a K-correction and combined with our inclination and mass ratio constraints would yield a precise NS mass measurement. Our measurements are consistent with \cite{Ozel2006}, who used PRE bursts to measure the mass and radius of EXO 0748$-$676 to be $M_{\rm ns} = 2.10 \pm 0.28 M_{\odot}$ and $r_{\rm ns} = 13.8 \pm 1.8$km respectively, thus ruling out soft EoS. Our findings improve confidence in their conclusion, agreeing that a harder EoS is required for nuclear matter.
% Moreover, the measured mass function is a {\it lower limit} on the true mass function, since it is inferred from emission lines that originate from the irradiated face of the companion star \citep{Bassa2009, Ratti2012}. The true mass function is calculated by applying a K-correction \citep{MunozDarias2005}; we have considered the most conservative (i.e. closest to unity) K-correction possible when implementing the radial velocity of H$_{\alpha}$, meaning that the mass is likely \textit{larger} than this measured value. When realistically implementing the Bowen blend's radial velocity measurement, the mass distributions can easily extend past $3 M_{\odot}$. Therefore, if we obtained direct measurements of emission lines from the companions surface or applied more precise K-corrections, the mass distributions will likely extend into the $3.0 - 5.0 M_{\odot}$ compact object mass-gap, thus future direct measurements of $K$ may even inform on the observational lower bound of the mass gap.
% Our results are also consistent with \citealt{MD2009} who performed emission line tomography to measure $M_\odot \leq M_{\rm ns} \leq 2.4 M_\odot$ and $0.11 \leq q \leq 0.28$.

\section{Conclusions}
\label{Section:Conclusion}
% The highly inclined neutron star low-mass X-ray binary, EXO 0748-676, experienced a $> 20$ year-long period of outburst, during which it was extensively monitored by \textit{EXOSAT}, \textit{XMM-Newton} and \textit{RXTE}, providing many clear and well-studied X-ray eclipses. In this paper we study 
% a binary mass function (Equation \ref{eqn:massfunc}) can be used to constrain the
% neutron star's 
We have studied archival \textit{XMM-Newton} observations of X-ray eclipses from EXO 0748$-$676 in the soft state, finding that they display a gradual decline in-to and out-of totality. Through timing analysis and phase-resolved spectroscopy, we uncover a narrow region of highly ionised material surrounding the companion star, which preferentially absorbs softer X-rays, creating energy-dependent eclipse profiles. The layer of material is found to be $\sim$ 20 - 40 per cent thicker behind the star than in the direction of orbital motion, thus explaining why the egress is observed to be $\sim 2.3$ s longer than the ingress. Similar (albeit more extreme) asymmetries are present in radio-band observations of eclipsing spider pulsars, as are frequency-dependent eclipses. Therefore, we favour an interpretation in which EXO 0748$-$676 is a spider pulsar progenitor. We invoke an \textit{early ablation} scenario, which introduces a small amount of absorbing material around the companion star. The material is suggested to originate from the surface of the companion star, having been evaporated off by incident radiation from the NS (and accretion flow) or perhaps a pulsar wind. Similar interpretations of EXO 0748$-$676 have been suggested by \citet{Parmar1991}, \citet{Ratti2012} and \citet{Parikh2021}.

We model the energy-dependent eclipse profiles of EXO 0748$-$676 to estimate the mass ratio, $q \sim 0.2$ and the binary inclination angle $i \sim 77 ^{\circ}$. In eclipsing systems, these values are related via the duration of totality, $t_{\rm e}$; therefore we can combine our measurement of $q$ with the previously measured binary mass function to constrain the NS mass. Using the DM derived semi-amplitude of the H$\alpha$ emission line, $K_{\rm em} \approx 410$ km/s \citep{Bassa2009}, yields $M_{\rm{ns}} \sim 2~M_{\odot}$ even for the most conservative (closest to unity) K-correction possible to account for the fact that this line originates from the irradiated face of the companion \citep{MunozDarias2005}. In this case, the canonical NS mass of $M_{\rm ns} = 1.4~M_\odot$ is outside of our $3 \sigma$ confidence contour, and use of a more realistic K-correction ($K_{\rm corr} \sim 0.8$) even pushes the posterior NS mass distribution into the $3 - 5~M_{\odot}$ mass gap. If we instead discard the H$_\alpha$ line and rely on the Bowen blend \citep{Mikles2012}, the canonical NS mass is permitted but the peak-posterior value for a realistic K-correction is still $M_{\rm{ns}} \sim 2~M_{\odot}$. Future observations of spectral lines emitted directly from the whole companion star surface could be combined with our inclination and mass ratio constraints to yield a precision NS mass measurement.

% We favour $M_{\rm{ns}} \sim 2.0 M_{\odot}$ using the semi-amplitude of the H$\alpha$ emission line, $K_{\rm em} \approx 410$ km/s \citep{Bassa2009}, and the most conservative K-correction possible to account for the fact that this line originates from the irradiated face of the companion \citep{MunozDarias2005}. \edit{In this case, the canonical NS mass of $M_{\rm ns} = 1.4~M_\odot$ is outside of our $3 \sigma$ confidence contour. Consistent NS masses ($M_{\rm{ns}} \sim 2.0 M_{\odot}$) are obtained when implementing the radial velocity measurement of $K_{\rm em} \approx 329$ km/s, constrained from the Bowen blend \citep{Mikles2012}, although these distributions are broader and do not rule out the canonical mass.}
% Our inferred masses indicate harder EoS are required for nucleonic matter. A more realistic K-correction could increase our inferred NS mass and the corresponding mass distributions would likely extend into the $3.0 - 5.0 M_{\odot}$ mass gap. Thus, future measurements of $K$ could inform the observational lower bound of the mass gap.
% We apply a conservative K-correction following \citet{MunozDarias2005}, using $K_{\rm{em}} \approx 410$ \citep{Bassa2009}, finding $M_{\rm{ns}} \sim 2.0 M_{\odot}$. Our measurements are consistent with previous mass measurements \citep{MD2009, Ozel2006}, improving confidence in their conclusions and indicating that harder EoS are required for nuclear matter. Our conservative approach to the K-correction yields a lower limit on $M_{\rm{ns}}$, meaning the mass is likely larger than our measurement.

\section*{Acknowledgements}
A. K. acknowledges support from the Oxford Hintze Centre for
Astrophysical Surveys, which is funded through generous support from the Hintze Family Charitable Foundation.
A. I. acknowledges support from the Royal Society.
The authors acknowledge helpful and insightful conversations with members of the accretion and a transient research group at the University of Oxford, led by Professor Rob Fender. In particular, we thank Jakob van den Eijnden for suggesting the accelerating wind density profile and Edward Nathan for his knowledge of running and testing the convergence of the Markov chains. 
We thanks the anonymous referee for insightful comments that improved the paper.

\section*{Data Availability}
The data used in this study are publicly available from the HEASARC website. The eclipse profile model is available upon reasonable request to the authors.

%%%%%%%%%%%%%%%%%%%%%%%%%%%%%%%%%%%%%%%%%%%%%%%%%%



\bibliographystyle{mnras}
\bibliography{biblio2} % if your bibtex file is called example.bib

% \begin{thebibliography}{99}
% \bibitem[\protect\citeauthoryear{Author}{2012}]{Author2012}
% Author A.~N., 2013, Journal of Improbable Astronomy, 1, 1
% \bibitem[\protect\citeauthoryear{Others}{2013}]{Others2013}
% Others S., 2012, Journal of Interesting Stuff, 17, 198
% \end{thebibliography}

%%%%%%%%%%%%%%%%%%%%%%%%%%%%%%%%%%%%%%%%%%%%%%%%%%

%%%%%%%%%%%%%%%%% APPENDICES %%%%%%%%%%%%%%%%%%%%%
\begin{appendices} 

\section{X-ray emitting region}
\label{Section:rx}

%Left, Bottom, Right, Top

If we assume that the companion star is an optically thick sphere with no surrounding material, the ingress duration is the time it takes for the outer radius of the companion star to cross the finite X-ray emitting region. In this scenario, we can calculate the radius of the (assumed spherical) X-ray emitting region by combining the ingress duration with the totality duration. Let us represent $\phi_1$ as the orbital phase at the start of the ingress, $\phi_2$ as the beginning of totality, $\phi_3$ the end of totality and $\phi_4$ the end of egress. Since the centre of totality is at $\phi=0$, we have that $\phi_1=-\phi_4$ and $\phi_2=-\phi_3$ where $\phi_3$ and $\phi_4$ are positive. The phase duration of totality is therefore $\Delta \phi_{\rm e} = 2\pi \Delta t_{\rm e}/P = \phi_3 - \phi_2 = 2\phi_3$, and the phase duration of ingress is $\Delta \phi_{\rm in} = 2\pi \Delta t_{\rm in}/P = \phi_2 - \phi_1= \phi_4-\phi_3$. The projected separation between the centre of the NS and the centre of the companion star at orbital phase $\phi$ is $R_{\rm cs} b(\phi)$, where the impact parameter $b(\phi)$ is given by Equation (\ref{eqn:b}). We can write $b(\phi_3)=1-r_x/R_{\rm cs}$ and $b(\phi_4)=1+r_x/R_{\rm cs}$ and solve for $\phi_3$ and $\phi_4$. After applying the small angle approximation ($\sin\phi \approx \phi$), we obtain
\begin{equation}
    \Delta \phi_e = \frac{2}{\sin i} \sqrt{ \left( \frac{R_{\rm cs} - r_x}{r_a} \right)^2 - \cos^2 i },
    \label{eqn:phie}
\end{equation}
and
\begin{equation}
    \Delta \phi_{\rm in} = \phi_2 - \phi_1 = \frac{1}{\sin i} \sqrt{ \left( \frac{R_{\rm cs} + r_x}{r_a} \right)^2 - \cos^2 i } - \frac{\Delta\phi_e}{2}.
    \label{eqn:phiin}
\end{equation}

Equations (\ref{eqn:phie}) and (\ref{eqn:phiin}) are a pair of simultaneous equations with three unknowns: $R_{\rm cs}/r_a$, $r_x/r_a$ and $i$. We can re-arrange these two equations to find
\begin{equation}
    2 \frac{r_x}{r_a} = \sqrt{ (\Delta\phi_{\rm in}+\Delta\phi_e/2)^2 \sin^2i + \cos^2i} - \sqrt{ (\Delta\phi_e/2)^2\sin^2i + \cos^2i}. 
\end{equation}
and
\begin{equation}
    2 \frac{R_{\rm cs}}{r_a} = \sqrt{ (\Delta\phi_{\rm in}+\Delta\phi_e/2)^2 \sin^2i + \cos^2i} + \sqrt{ (\Delta\phi_e/2)^2\sin^2i + \cos^2i}. 
\end{equation}

\begin{figure}
\centering
\includegraphics[width=\columnwidth,trim=2cm 1.5cm 2.5cm 3cm,clip=true]{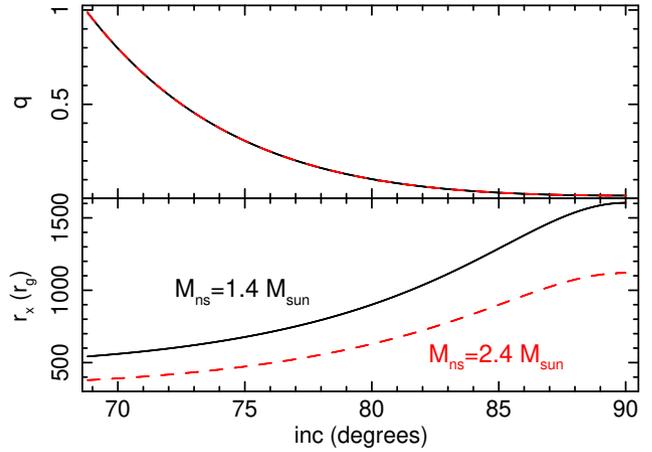}
\caption{Radius of the X-ray emitting region (bottom) and mass ratio (top) inferred from the ingress duration and totality duration assuming that the companion star is an optically thick sphere with no surrounding material. Black solid and red dashed lines correspond to an assumed NS mass of $1.4~M_\odot$ and $2.4~M_\odot$ respectively.}
\label{fig:rxrg}
\end{figure}

Therefore, for a given inclination angle, we can calculate $r_x/r_a$ and $R_{\rm cs}/r_a$. We can then assume that the companion fills its Roche-Lobe in order to calculate the mass ratio $q$ from $R_{\rm cs}/r_a$ (Equation \ref{eqn:RLfill}). In order to calculate $r_x$ as a function of $i$, we then only need to assume a NS mass and calculate $r_a$ from Kepler's law. 

\begin{figure*}
\centering
\includegraphics[width=\textwidth]{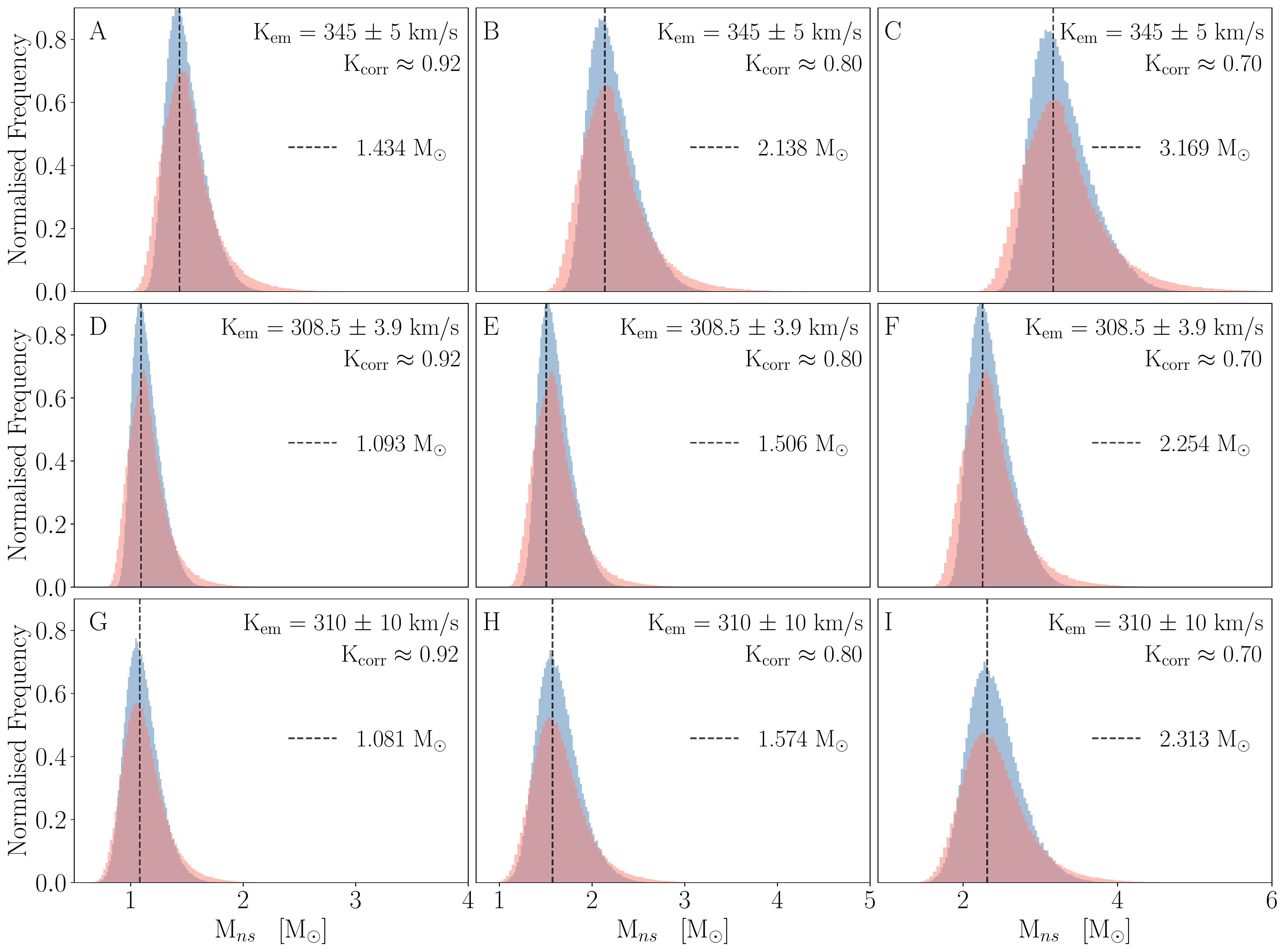}
\caption{\label{fig:KMC} Panels A-I: Distributions of $M_{\rm{ns}}$ assuming the Gaussian (blue) and exponential (red) density profiles within the eclipse profile model calculated using a different combination of $K_{\rm em}$ and K-Correction, which are detailed in each panel. Also shown is a black dashed line corresponding to the mean peak NS mass. The K-correction in A, D and G corresponds to the most conservative value calculated using the mass ratio dependent relation of \citet{MunozDarias2005} (Equation \ref{eqn:kcorr}) for our best fitting value $q = 0.222$. Column 2 (B, E and H) and column 3 (C, F and I) assume the reasonable K-corrections of 0.8 and 0.7 respectively. Emission lines considered are 345 $\pm$ 5 km/s from He$_{\rm{II}}$ \citep{Bassa2009}, 310 $\pm$ 10 km/s from He$_{\rm{II}}$ \citep{MD2009} and 308.5 $\pm$ 3.9 km/s from a weighted average of H$_{\beta}$ and H$_{\gamma}$ \citep{Ratti2012}. }
\end{figure*}

Figure \ref{fig:rxrg} shows the resulting inferred values of $r_x$ and $q$ as a function of inclination angle for two values of NS mass. We see that $r_x$ increases with $i$, and recover the simple equation $r_x(i=90^\circ) = \Delta \phi_{\rm in} r_a / 2$ from the main text. For inclination lower than $\sim 69^\circ$, there is no solution for $q$. The smallest possible X-ray region under our assumptions is therefore $r_x\sim 400~r_g$. This is implausibly large, and adds to the argument that the ingress duration must instead be dominated by an extended stellar atmosphere and/or a layer or ablated material.

\begin{figure*}
\centering
\includegraphics[width=\textwidth]{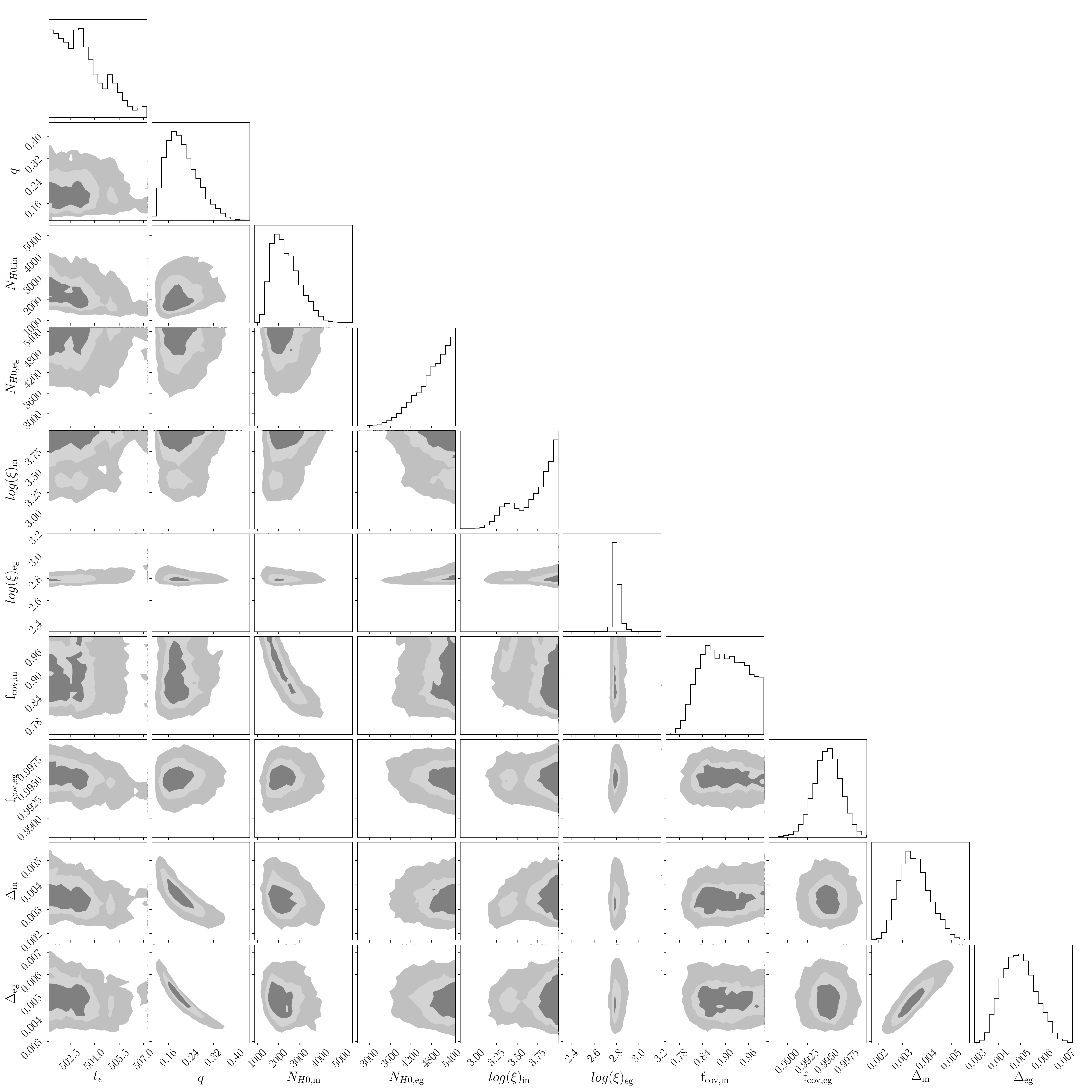}
\caption{Output distributions from the MCMC simulation of the eclipse profile model with the Gaussian density profile with a chain length of 307200, a burn-in period of 92160 and 256 walkers. Hard upper limits of 5500 are used for the surface column densities, $N_{H0,\rm{in}}$ and $N_{H0,\rm{eg}}$ to prevent the walkers entering non-physical parts of parameters space. For the same reason the ionisation parameters, $\log(\xi)_{\rm{in}}$ and $\log(\xi)_{\rm{eg}}$, had an upper limit of 4.0. The lines and shading, dark to light, on the 2D histograms represent $1 \sigma$, $2 \sigma$ and $3 \sigma$ contours respectively. The y-axes for the 1D histograms are in arbitrary units.}
\label{fig:cornerGA}
\end{figure*}

\begin{figure*}
\centering
\includegraphics[width=\textwidth]{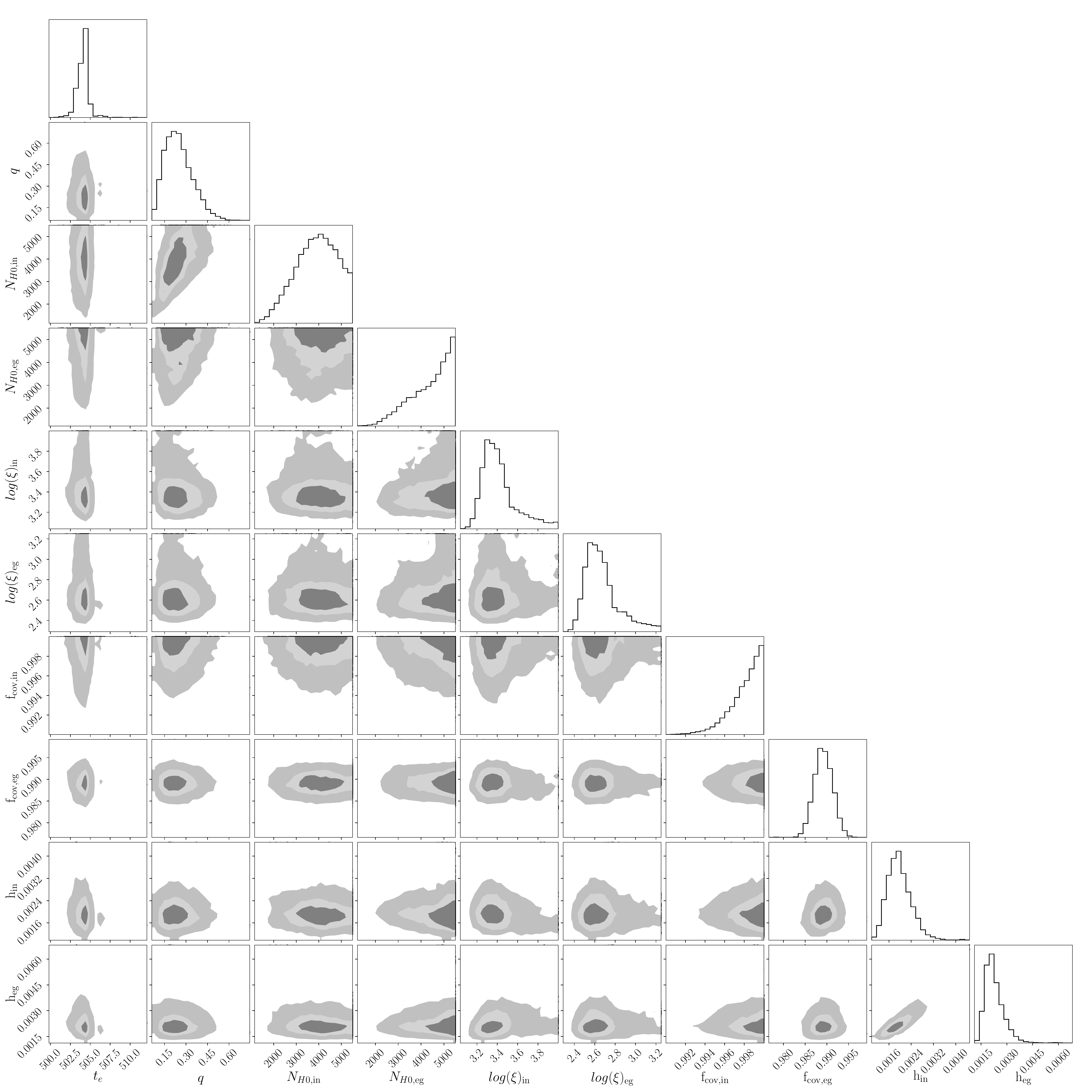}
\caption{Output distributions from the MCMC simulation of the eclipse profile model with the exponential density profile with a chain length of 307200, a burn-in period of 92160 and 256 walkers. Hard upper limits of 5500 are used for the surface column densities, $N_{H0,\rm{in}}$ and $N_{H0,\rm{eg}}$ to prevent the walkers entering non-physical parts of parameters space. For the same reason the ionisation parameters, $\log(\xi)_{\rm{in}}$ and $\log(\xi)_{\rm{eg}}$, had an upper limit of 4.0. The lines and shading, dark to light, on the 2D histograms represent $1 \sigma$, $2 \sigma$ and $3 \sigma$ contours respectively. The y-axes for the 1D histograms are in arbitrary units.}
\label{fig:cornerEx}
\end{figure*}

\section{Neutron Star Mass}
\label{Section:Mass}
Figure \ref{fig:KMC} is provided to show the posterior NS mass distributions obtained using $K_{\rm{em}}$ values from \citealt{Bassa2009}, \citealt{Ratti2012} and \citealt{MD2009}. Panels A, B and C assume $K_{\rm{em}} = 345 \pm 5$ km/s \citep{Bassa2009}. This value is subject to the same uncertainties as $K_{\rm{em}} = 410 \pm 5$ km/s \citep{Bassa2009}, which is presented in the main text and therefore, is not included in our primary analysis. Panels D, E and F assume $K_{\rm{em}} = 308.5 \pm 3.9$ km/s \citep{Ratti2012}. Panels G, H and I assume $K_{\rm{em}} = 310 \pm 10$ km/s \citep{MD2009}. The latter two radial velocity measurements were not considered in the main text as they are lower than other measured radial velocities, indicating that they originated from a spatial zone further from the centre of mass of the companion and are, therefore, less constraining.

\section{Markov Chain Monte Carlo}
\label{Section:ApMCMC}

To further understand the parameter space of the eclipse mapping model, we run a Markov Chain Monte Carlo (MCMC) simulation within \texttt{xspec} using the Goodman-Weare algorithm. We run 4 chains using the intrinsic routine \texttt{chain}, two assuming the Gaussian density profile and two assuming the exponential density profile. Each chain is run individually, ensuring that the chains assuming the same radial density profile are not correlated. We use a chain length of 307200 with a burn-in period of 92160. Each chain uses 256 walkers and starts, respectively, from their best fits presented in Table \ref{tb:fitpars}. 

Our model assumes a constant out-of-eclipse spectrum so only parameters governing the eclipse profiles are variable during the fits. Therefore, we have 12 free parameters. Two of the parameters are simply normalise of the eclipse profiles such that the out-of-eclipse count rate equals 1.0 and the time at the centre of totality equals 0.0, so there are 10 key parameters we explore here: $t_e$, $q$, $N_{H0, \rm{in}}$, $N_{H0, \rm{eg}}$, $\log(\xi)_{\rm{in}}$, $\log(\xi)_{\rm{eg}}$ $\rm{f_{cov, in}}$, $\rm{f_{cov, in}}$, $\Delta_{\rm{in}}$, $\Delta_{\rm{eg}}$ (for the Gaussian density profile only), $\rm{h_{in}}$ and $\rm{h_{eg}}$ (exponential density profile only). Figures \ref{fig:cornerGA} and \ref{fig:cornerEx} show the output distributions for the Gaussian and exponential models respectively.

For the eclipse profile model with a Gaussian radial density profile, we see evidence of a positive correlation between $\Delta_{\rm{in}}$ and$\Delta_{\rm{eg}}$. Assuming a spherically symmetric system, it can be expected that the amount of material accumulating around the star to increase similarly on both sides. The mass ratio is anti-correlated with both $\Delta_{\rm{in}}$ and$\Delta_{\rm{eg}}$ which is expected from our formalism presented in Sections \ref{Section:TRS} and \ref{Section:ECProfs}. Similarly, the scale heights, $\rm{h_{in}}$ and $\rm{h_{eg}}$ in the eclipse profile model with an exponential radial density profile display a slightly positive correlation. Similar physical arguments to those presented above can explain this relationship. No other correlations are present for these models. 

The convergence of each chain is tested using the Geweke convergence measure which compares the mean of each parameter in two intervals of the chain, one shortly after the burn-in period and one towards the end of the chain. These correspond to the first $10 \%$ and the last $50 \%$ of the chain. For all chains, all parameters measured between $\pm 0.2$, indicating that convergence has been achieved. One parameter remains an exception to this: $N_{H0, \rm{eg}}$ in the eclipse profile model assuming an exponential density profile. This parameter measures as 0.545 and 0.551 for the two chains. The higher Geweke values here are likely a result of an upper limit on the parameter of 5500, which is in place to prevent the walkers from venturing into nonphysical parts of parameter space. Therefore, we remain confident in the convergence of the chain despite this result.

To increase our confidence in the convergence of the chains, we use the Rubin-Gelman convergence test to compare the variances of each parameter in the chain between two chains of the same length. For this test, we implement a stringent bound of $R_c < 1.1$ as an indication of convergence, with higher values indicating parameters are varying too much and have not yet converged. For both models, all parameters give $R_c < 1.05$, therefore we are confident in the convergence of the chains.

\end{appendices}

%%%%%%%%%%%%%%%%%%%%%%%%%%%%%%%%%%%%%%%%%%%%%%%%%%

% Don't change these lines
\bsp	% typesetting comment
\label{lastpage}
\end{document}